\documentclass[a4paper,11pt]{article}
\usepackage{marvosym}
\usepackage{comment}
\usepackage{subfigure}
\usepackage{mathrsfs} 
\usepackage[framemethod=default]{mdframed}
\newmdenv[skipabove=1pt,
skipbelow=1pt,
rightline=false,
leftline=false,
topline=false,
bottomline=false,
backgroundcolor=gray!10,
linecolor=gray,
innerleftmargin=5pt,
innerrightmargin=5pt,
innertopmargin=-10pt,
innerbottommargin=5pt,
leftmargin=0cm,
rightmargin=0cm,
linewidth=4pt]{eBox}

\newcommand{\Di}[1]{\prod_{i=1}^{#1} \mathfrak{D}^{M_{i} A_{i}} }
\newcommand{\V}{{\mathbf{{V}}}}

\catcode`\,12

\pdfoutput=1 

\usepackage{jheppub} 
\usepackage[T1]{fontenc} 

\title{Towards the Feynman rule for $n$-point gluon Mellin amplitudes in AdS/CFT}

\author{Jinwei Chu,}
\author{Savan Kharel}

\affiliation{Department of Physics, University of Chicago, Chicago, IL 60637, USA}

\emailAdd{jinweichu@uchicago.edu}
\emailAdd{skharel@uchicago.edu}

\abstract{We investigate the embedding formalism in conjunction with the Mellin transform to determine tree-level gluon amplitudes in AdS/CFT. Detailed computations of three to five-point correlators are conducted, ultimately distilling what were previously complex results for five-point correlators into a more succinct and comprehensible form. We then proceed to derive a recursion relation applicable to a specific class of $n$-point gluon amplitudes. This relation is instrumental in systematically constructing amplitudes for a range of topologies. We illustrate its efficacy by specifically computing six to eight-point functions.
Despite the complexity encountered in the intermediate steps of the recursion, the higher-point correlator is succinctly expressed as a polynomial in boundary coordinates, upon which a specific differential operator acts. Remarkably, we observe that these amplitudes strikingly mirror their counterparts in flat space, traditionally computed using standard Feynman rules. This intriguing similarity has led us to propose a novel dictionary: comprehensive rules that bridge AdS Mellin amplitudes with flat-space gluon amplitudes.
}

\usepackage{amsmath,amssymb,amsfonts,mathtools,bm} 
\numberwithin{equation}{section}
\usepackage{physics} 
\usepackage[margin=10pt,font=small,labelfont=bf]{caption}
\usepackage[pdftex,dvipsnames]{xcolor}
\definecolor{darkblue}{rgb}{0.1,0.1,.7}
\usepackage{graphicx}  
\graphicspath{ {images/} }
\mathcode`l="8000
\begingroup
\makeatletter
\lccode`\~=`\l
\DeclareMathSymbol{\lsb@l}{\mathalpha}{letters}{`l}
\lowercase{\gdef~{\ifnum\the\mathgroup=\m@ne \ell \else \lsb@l \fi}}%
\endgroup
\def\<#1\>{\expval{#1}}

\newcommand{\undertl}[1]{\mathord{\vtop{\ialign{##\crcr
				$\hfil\displaystyle{#1}\hfil$\crcr\noalign{\kern1.5pt\nointerlineskip}
				$\hfil\tilde{}\hfil$\crcr\noalign{\kern1.5pt}}}}} 

\newcommand   \lptl{\raise .8ex\hbox{$^\leftarrow$} \hspace{-9pt} \partial} 
\newcommand   \lrptl{\raise .8ex\hbox{$^\leftrightarrow$} \hspace{-9pt} \partial} 
\def\be#1\ee{\begin{equation}\begin{aligned}#1\end{aligned}\end{equation}}
\makeatletter
\DeclareRobustCommand\bea{\@ifnextchar[{\@@bea}{\@bea}}
\def\@@bea[#1]#2\eea{\begin{subequations}\begin{align}#2\end{align}\label{#1}\end{subequations}}
\def\@bea#1\eea{\begin{subequations}\begin{align}#1\end{align}\end{subequations}}
\makeatother
 \catcode`,\active
 
 \catcode`\,12
 
\renewcommand \l  {\lambda}

\begin{document} 
\maketitle
\flushbottom

\section{Introduction}
In recent decades, the study of holographic theories has become a significant area of theoretical research. Among these theories, the most developed are those within the framework of asymptotically Anti-de Sitter (AdS) spacetimes \cite{Maldacena:1997re, Witten:1998qj}. These theories present a contrast to the traditional `in' and `out' states found in Minkowski spacetime, crucial for scattering amplitudes. In AdS spacetimes, particles are intrinsically confined, leading to perpetual interactions. Despite this, interactions at the timelike boundary of AdS permit the creation and annihilation of particles within this spacetime. Notably, these transition amplitudes in AdS have a direct analogy to correlation functions in the corresponding Conformal Field Theory (CFT). This correlation allows for the interpretation of CFT correlation functions as scattering amplitudes in the AdS context.

Scattering amplitudes in Anti-de Sitter (AdS) space can be computed using Witten diagrams, which are the AdS counterparts of Feynman diagrams used in flat space. 
Initial efforts by researchers to extend their analysis beyond three or four-point Witten diagrams faced significant computational challenges (e.g. \cite{Liu:1998ty,Freedman:1998bj,DHoker:1998bqu,Liu:1998th,DHoker:1998ecp}). Two primary difficulties emerged in this field: the complexity of bulk integrals and the intricacies involved in dealing with spin-bearing external operators. Overcoming these challenges has defined much of the ongoing research in this area. In this paper, we directly address both these challenges. We will embark on the calculation of higher-point external spinning field.

Our work draws inspiration from the recent wave of diverse and intriguing contributions to the computation of both scalar and spinning correlators in AdS, employing varied methodologies: momentum space \cite{Raju:2012zr, Albayrak:2018tam, Albayrak:2019yve,Albayrak:2019asr, Albayrak:2020bso,Albayrak:2020isk, Albayrak:2020fyp, Albayrak:2023kfk, Bzowski:2013sza,Bzowski:2015pba, Bzowski:2018fql,Bzowski:2019kwd, Bzowski:2020kfw,  Bzowski:2022rlz,Bzowski:2023jwt,Jain:2020rmw, Jain:2021vrv,  Jain:2023idr, Isono:2018rrb, Isono:2019wex, Coriano:2013jba, Coriano:2018bbe, Anand:2019lkt,  Caloro:2022zuy, Meltzer:2021zin}, position-Mellin space 
\cite{Mack:2009gy,Mack:2009mi,Penedones:2010ue,Paulos:2011ie,Fitzpatrick:2011ia,Kharel:2013mka,Penedones:2019tng,Rastelli:2016nze,Zhou:2017zaw,Rastelli:2017udc,Alday:2020dtb,Alday:2020lbp,Alday:2021ajh, Gopakumar:2016cpb, Jepsen:2018dqp, Jepsen:2018dqp, Jepsen:2019svc,Li:2023azu, Gadde:2022ghy, Bianchi:2021piu, Hoback:2020pgj,Haldar:2019prg}, and more recently momentum-Mellin space~\cite{Sleight:2021iix, Sleight:2019hfp,Mei:2023jkb}.
In this paper, our primary focus is on studying gluon scattering within AdS. We find the advancements in flat space scattering amplitudes, particularly those involving gluon and graviton scattering, to be remarkably intriguing ~\cite{Elvang:2013cua}. These developments not only bolster experimental results at major colliders like the LHC but also revitalize foundational quantum field theory research. A notable aspect of gluon amplitudes in flat space is their simplicity and elegance; despite the complexity of intermediate calculations, the final results, as epitomized by the classic Parke-Taylor formula, are often concise and elegant \cite{Parke:1986gb}. Furthermore, these advancements unveil fascinating connections between core physics and diverse mathematical fields~\cite{Arkani-Hamed:2013jha,Arkani-Hamed:2017mur}. 
Motivated by these developments, our focus is on studying gluon scattering within AdS.

We use Mellin space as our investigative tool and it offers unique advantages. In Mellin space, amplitudes are clearly presented as meromorphic functions of their variables, echoing the well-understood analytic properties of the S-Matrix in flat space. However, Mellin space has not been fully explored, especially when examining spinning correlators \cite{Goncalves:2014rfa, Costa:2014kfa, Nishida:2018opl, Sleight:2017fpc}. Our study additionally focuses on addressing the challenging issue of higher-point correlators with external spin, a task underscored by the limited amount of analytical work in this area due to its technical complexity. Yet, these higher-point analysis are important for major theoretical breakthroughs. Insights from the modern S-matrix program show that deeper exploration of higher-point gauge and gravity amplitude (including loop amplitudes) has greatly helped us unravel deep mathematical structures. Hence, a thorough examination of higher-point spinning structures in Anti-de-Sitter (AdS) space is essential to uncover potential simplicities and mathematical insights akin to flat space scattering amplitudes.

In addition to their relevance in Anti-de-Sitter (AdS) space, these structures carry broader implications. Notably, they are interconnected with de Sitter (dS)  \cite{Maldacena:2002vr,Ghosh:2014kba} aligning well with the program to construct cosmologically relevant correlators \cite{Arkani-Hamed:2018kmz,Baumann:2020dch,Sleight:2020obc,Arkani-Hamed:2023kig}. Spinning correlators in AdS could have substantial importance in the cosmological frontier. Moreover, specific case studies are crucial for advancing our understanding of the still-ambiguous double copy principle in curved spacetime. This principle is particularly important when applied to higher-point structures, and thus, concrete examples are indispensable for its possible formulation akin to flat space. 

In this paper, we unveil a formalism anchored in embedding-space techniques to meet our research objectives. Utilising key differential operators, we streamline the complex calculations tied to higher-point correlators with external spinning fields. By methodically building upon lower-point AdS correlators, we achieve recursive computations of higher-point amplitudes in AdS. The paper's structure is as follows: In Section \ref{Section2}, we articulate the foundational principles and techniques vital for AdS amplitude calculations. We delve into the embedding formalism specific to AdS space and highlight the role of Mellin space as an eigenspace for these amplitudes. We also present a summary of our main results. Section \ref{Section3} offers a comprehensive computation of three, four, and five-point amplitudes, paving the way for subsequent, more nuanced higher-point analysis. Here the elegant mapping between flat-space Feynman rules and AdS begins to emerge. In Section \ref{Section4}, we derive a recursion formula for $n$-point amplitudes, to assist an ambitious calculation of six-point, seven-point, and eight-point gluon topologies. Notably, we again notice that Mellin amplitudes for gluons strikingly parallel flat-space scattering amplitudes, despite the complexity of intermediate calculations. This revelation leads us to propose a remarkably streamlined map to flat space for $n$-point gluon amplitudes. Finally, we discuss important work that can spur from our results in Section \ref{section5}. 

This paper is a substantial expansion of the companion version~\cite{Chu:2023pea} which we recommend to the reader who want to skip technical details and interested in the main essence on the first reading.

\section{Preliminaries and summary}
\label{Section2}
AdS amplitude is holographically dual to Conformal Field Theory correlation function, $\langle \mathcal{O}_1(P_1) \cdots \mathcal{O}_n(P_n) \rangle$ where $P_i$ denotes the AdS boundary coordinate where the operator $\mathcal{O}_i$ is inserted. Here we provide an overview of the fundamental ingredients and concepts involved in calculating AdS amplitudes.
\subsection{Embedding space formalism}

The calculation of Witten diagrams is markedly streamlined with the application of the embedding formalism \cite{Penedones:2010ue}.\footnote{In a seminal work by Dirac \cite{Dirac:1936fq}, it was proposed that the conformal group $SO(d + 1, 1)$ naturally ``lives'' in the embedding space \(\mathbb{R}^{d+1,1}\). Here, it can be understood as the group of linear isometries. This suggests that constraints imposed by conformal symmetry could be as straightforward as those from Lorentz symmetry. Also see Weinberg's paper \cite{Weinberg:2010fx}.} This formalism stands as a robust tool for the in-depth exploration and analysis of the properties and dynamics inherent in AdS spaces. This formalism allows us to describe an AdS$_{d+1}$ space by embedding it in a higher-dimensional Minkowski space, denoted as $\mathbb{R}^{d+1,1}$. AdS coordinate vectors $X$ satisfy the following property,
\be
\label{AdS}
X\cdot X\equiv \eta_{MN}X^MX^N=-R^2 .
\ee
Throughout the paper we will take $R=1$. The boundary of the AdS$_{d+1}$ space is at $X\to\infty$, where (\ref{AdS}) asymptotes to an equation of a light cone. It is convenient to think of the conformal boundary of AdS as the space of null rays. 

We use $P$ to denote the fixed boundary point. Hence, $P\cdot P\equiv \eta_{MN}P^MP^N=0$.
Therefore, the distance between any two boundary points $P_i$ and $P_j$ is defined by $P_{ij}\equiv \left(P_i-P_j\right)^2=-2P_i\cdot P_j$.

\subsection{Mellin space}
Another key mathematical apparatus utilized in our study is the Mellin space.\footnote{See \cite{Fitzpatrick:2011ia} for a nice review of Mellin Space in the AdS/CFT.}  Mellin amplitudes have structural similarity to flat space momentum space scattering amplitudes. Many researchers have demonstrated that the Mellin representation has advantages in analyzing CFT correlation functions, particularly within the large $N$ expansion. 

The basis of Mellin space is $\prod_{i<j}P_{ij}^{-\gamma_{ij}}$, where $\gamma_{ij}$ are called Mellin variables. The scaling dimension of this basis for $P_i$ is $\sum_{j\neq i}\gamma_{ij}$. First, we focus on the scalar cases. Expanded in Mellin space, an $n$-point amplitude can be expressed as
\be
\label{Mellin}
\left\langle \prod_{i=1}^n\mathcal{O}_i(P_i)\right\rangle=\int  \left(\prod_{i< j}^n\frac{d\gamma_{ij}}{2\pi i}\Gamma(\gamma_{ij})P_{ij}^{-\gamma_{ij}}\right)\prod_{i=1}^n\delta\Big(\sum_{j\neq i}\gamma_{ij}-\Delta_i\Big)\mathscr{M}_n(\gamma_{ij})\ ,
\ee
where $\mathscr{M}_n(\gamma_{ij})$ is called Mellin amplitude. Note that the delta functions restrict the correct scaling behavior of $\mathcal{O}_i(P_i)$. For the sake of notational simplicity, we will forgo including them in our subsequent equations. 

In the context of vector fields $J^{M_i}(P_i)$, our primary interest in this paper, the amplitude takes on a slightly different form to incorporate the indices. We can write it as
\be
\label{vecMellin}
\left\langle \prod_{i=1}^nJ^{M_i}(P_i)\right\rangle=\int  \left(\prod_{i< j}^n\frac{d\gamma_{ij}}{2\pi i}\Gamma(\gamma_{ij})P_{ij}^{-\gamma_{ij}}\right)\mathscr{M}_n^{M_1M_2\cdots M_n}(\gamma_{ij},P_i)\ .
\ee
In this context, it is crucial to underline a subtle difference as compared to the scalar scenario. Specifically, the Mellin amplitude \( \mathscr{M}_n^{M_1M_2\cdots M_n}(\gamma_{ij}, P_i) \) is a function not only of the Mellin variables \( \gamma_{ij} \), but also of the boundary coordinates \( P_i \). This is attributed to the possibility that vector Mellin amplitude may contain \( P_i \) with free indices.\footnote{More generally, each field in the correlation function has a spin of $l_i$. Then, there are totally $\sum_{i=1}^nl_i$ free indices in the Mellin amplitude.}

\subsection{AdS amplitudes and toolkit}

Witten diagram, a powerful tool for computing amplitudes in Anti-de Sitter space, provides a systematic approach to analyze scattering processes. It is composed of two key elements: vertices and propagators. Vertices represent the interaction points where particles or fields within the AdS theory come together. They are integrated over the entire AdS space, encapsulating the bulk interactions.

Propagators, on the other hand, come in two forms: Boundary-to-bulk propagators connect a point on the AdS boundary to a vertex in the bulk, capturing the information flow from the boundary into the bulk. Meanwhile, bulk-to-bulk propagators link two vertices within the bulk, accounting for the propagation of particles or fields between these interaction points.

\subsubsection*{Scalar}
The boundary-to-bulk propagator for a scalar field $\mathcal{O}_i$ is a function of the boundary point $P_i$ and the bulk point $X$, i.e.,
\be
\label{EPiX}
\mathcal{E}(P_i,X)=\frac{C_{\Delta_i}}{(-2P_i\cdot X)^{\Delta_i}}\ ,\quad C_{\Delta_i}=\frac{\Gamma (\Delta_i)}{2\pi^h\Gamma \left(\Delta_i+1-h\right)}\ ,
\ee
where $h\equiv d/2$. To illustrate this, let's consider the calculation of the three-point scalar amplitude. In this case, we can compute the amplitude by utilizing the boundary-to-bulk propagator in the following straightforward manner:
\be
\left\langle \mathcal{O}_1(P_1)\mathcal{O}_2(P_2)\mathcal{O}_3(P_3)\right\rangle=ig \int_{\text{AdS}}dX\ \mathcal{E}(P_1,X)~ \mathcal{E}(P_2,X) \mathcal{E}(P_3,X)\ ,
\ee
where $g$ is the coupling constant. The Mellin amplitude, as it turns out (see Appendix \ref{3ptappendix} for more details), is given (as shown in for instance  \cite{Paulos:2011ie})
,\be
\label{3ptscalar}
\mathscr{M}_3(P_1,P_2,P_3)=ig\frac{\pi^h}{2}\prod_{i=1}^3\frac{C_{\Delta_i}}{\Gamma(\Delta_i)}\Gamma\left(\frac{\Delta_1+\Delta_2+\Delta_3-d}{2}\right)\ .
\ee
\subsubsection*{Vector}
In this paper, we compute higher-point amplitudes, taking into account fields with spinning degrees of freedom in both the internal propagator and external state. The boundary-to-bulk propagator for a vector field can be obtained by applying a differential operator to a scalar boundary-to-bulk propagator~\cite{Paulos:2011ie}. These operators act as projectors, projecting the spinning Mellin amplitude $\mathscr{M}_n^{M_1M_2\cdots M_n}$ onto a subspace that remains conformally invariant. Specifically, for a vector field \(J^{M_i}(P_i)\),
\begin{equation}
 \mathcal{E}^{M_iA_i}(P_i,X)=\widehat{D}^{M_iA_i} \mathcal{E}(P_i,X),
\end{equation}
where the operator \(\widehat{D}^{M_iA_i}\) is defined as follows:
\begin{equation}
\label{DMA}
\widehat{D}^{M_iA_i}=\frac{\Delta_i-1}{\Delta_i}\eta^{M_iA_i}+\frac{1}{\Delta_i}\frac{\partial}{\partial P_i^{M_i}}P_i^{A_i}\ .
\end{equation}

We want to highlight that the operator $\widehat{D}^{M_iA_i}$ simplifies the index structure of vector amplitudes, making it easier to relate to scalar amplitudes. In anticipation of future computations and for the sake of notational simplicity, let us introduce a concise version of the operator as follows:
\be
\left( \Di{n} \right) =\prod_{i=1}^n \frac{C_{\Delta_i}}{\Gamma(\Delta_i)} \widehat{D}^{M_iA_i}\ .
\ee
\subsubsection*{\'Etude of momentum conservation analogues}
We provide some properties of the differential operator given in (\ref{DMA}). This observation will be instrumental in deriving analogues of momentum conservation, as illustrated below.

An eigenfunction of the differential operator $\widehat{D}^{M_iA_i}$ can be expressed as $\frac{\partial}{\partial P_i^{A_i}}F_{\delta_i}(P_i)$, where $F_{\delta_i}(P_i)$ denotes any function of $P_i$ with the scaling dimension of $\delta_i$. That is, $P_i\cdot \frac{\partial}{\partial P_i}F_{\delta_i}(P_i)=-\delta_i F_{\delta_i}(P_i)$. Then,
\be
\label{eq:eigD}
\widehat{D}^{M_iA_i}\frac{\partial}{\partial P_i^{A_i}}F_{\delta_i}(P_i)=\frac{\Delta_i-1-\delta_i}{\Delta_i}\frac{\partial}{\partial P_i^{M_i}}F_{\delta_i}(P_i)\ .
\ee
Notably, when $\delta_i=\Delta_i-1$, or the scaling dimension of the eigenfunction $\frac{\partial}{\partial P_i^{A_i}}F_{\delta_i}(P_i)$ is $\Delta_i$, the eigenvalue is zero. 

Let's see couple of examples. Firstly, by substituting $F_{\Delta_i-1}=f(\gamma_{ij})\prod_{l<m}\Gamma(\gamma_{lm})P_{lm}^{-\gamma_{lm}}$ for some $i$ (with any function $f(\gamma_{ij})$ of Mellin variables $\gamma_{ij}$ for all $j\neq i$) in (\ref{eq:eigD}), with $\Delta_i-1=\sum_{j\neq i}\gamma_{ij}$, we deduce that
\begin{multline}
    \label{eigD1}
    0 =\int \prod_{l<m}d\gamma_{lm}\ \widehat{D}^{M_iA_i}\sum_{k\neq i}\big[P_{k,A_i}f(\gamma_{ij})\Gamma(\gamma_{ik}+1)P_{ik}^{-\gamma_{ik}-1}\prod_{\substack{l<m \\ (lm)\neq (ik)}}\Gamma(\gamma_{lm})P_{lm}^{-\gamma_{lm}}\big] \\
    = \int \prod_{l<m}d\gamma_{lm}\ \widehat{D}^{M_iA_i}\sum_{k\neq i}\big[P_{k,A_i}f(\gamma_{ij,j\neq k},\gamma_{ik}-1)\prod_{l<m}\Gamma(\gamma_{lm})P_{lm}^{-\gamma_{lm}}\big].
\end{multline}
In the final step we have shifted the Mellin variables, $\gamma_{ik}\to\gamma_{ik}-1$.\footnote{So now $\sum_{j\neq i}\gamma_{ij}=\Delta_i$.}

As another example, by substituting $F_{\Delta_{i_1}-1}=P_{i_1,A_{i_2}}f(\gamma_{i_1j})\prod_{l<m}\Gamma(\gamma_{lm})P_{lm}^{-\gamma_{lm}}$, for some $i_1$ and $i_2$, into (\ref{eq:eigD}), we deduce that
\begin{multline}
\label{eigD2}
0=\int \prod_{l<m}d\gamma_{lm}\ \widehat{D}^{M_{i_1}A_{i_1}}\big[\eta_{A_{i_1}A_{i_2}}f(\gamma_{i_1j})\prod_{l<m}\Gamma(\gamma_{lm})P_{lm}^{-\gamma_{lm}}\\
-2P_{i_1,A_{i_2}}\sum_{k\neq i_1}P_{k,A_{i_1}}f(\gamma_{i_1j,j\neq k},\gamma_{i_1k}-1)\prod_{l<m}\Gamma(\gamma_{lm})P_{lm}^{-\gamma_{lm}}\big]\ .
\end{multline}

These identities are crucial for significantly simplifying our target expression for higher-point functions and uncovering underlying structures.

\subsubsection*{Bulk-to-bulk propagators}
A bulk-to-bulk propagator represents the exchange of a primary field with a scaling dimension of $\Delta$, including its descendant fields. It is convenient that such propagators for both scalar and spinning particles can be written as the product of two boundary-to-bulk propagators glued together by integration over the boundary point $Q$. This property can help us recycle the lower-point function and obtain the higher-point function by appropriately gluing lower-point amplitudes. For pedagogical value, we first write the propagator associated with simpler scalar fields,
\be
\label{bbpsc}
\mathcal{G}_\Delta(X_1,X_2)=\int_{-i\infty}^{i\infty}\frac{dc}{2\pi i}\frac{2c^2}{c^2-(\Delta-h)^2}\int_{\partial\text{AdS}} dQ\ \mathcal{E}_{h+c}(Q,X_1)\mathcal{E}_{h-c}(Q,X_2)\ .
\ee
We can deform the integration contour in (\ref{bbpsc}) and integrate around the pole, e.g., $c=\Delta-h$. We subsequently get
\be
\mathcal{G}_\Delta(X_1,X_2)=\left(h-\Delta\right)\int_{\partial\text{AdS}} dQ\ \mathcal{E}_{\Delta}(Q,X_1)\mathcal{E}_{d-\Delta}(Q,X_2)\ .
\ee
Similarly, for vector fields, the bulk-to-bulk propagator is~\cite{Paulos:2011ie}
\be
\label{eq:bbv}
\mathcal{G}^{AB}_\Delta (X_1,X_2)=\int_{-i\infty}^{i\infty}\frac{dc}{2\pi i}f_\Delta(c)\int_{\partial\text{AdS}} dQ\  \mathcal{E}^{MA}_{h+c}(Q,X_1)\eta_{MN} \mathcal{E}^{NB}_{h-c}(Q,X_2)\ ,
\ee
where
\be
f_\Delta(c)=\frac{4c^2(h^2-c^2)}{\left(c^2-(\Delta-h)^2\right)^2}\ .
\ee

In principle, the existence of second-order poles in $f_\Delta(c)$ complicates the calculation of the contour integral. However, we will show that for a bulk-to-bulk propagator (one end of which is a three-vertex connected to two external fields on the boundary), these second-order poles simplify to first-order poles! This simplification enables easier integration with respect to $c$ and facilitates a recursive calculation of amplitudes in the channel with at most a single four-vertex. In summary, the structure of vector amplitudes is similar to that of scalar fields, i.e. both can be decomposed into products of lower-point amplitudes.
\subsection{Summary of the main results}
In this paper, we explicitly calculate the gluon Mellin amplitudes for several diagrams, spanning from three points to eight points. In addition to detailed calculations, this paper also serves as a repository for explicit higher-point results. To assist the reader, here we direct the reader to the main results of the paper.

The three-point gluon Mellin amplitude is presented in \eqref{eq:3J1}. Similarly, the four-point amplitudes include both contact and exchange channels. The explicit results for these are given respectively in \eqref{eq:4Jcontact1} and \eqref{eq:M_s}, where we have reproduced the results calculated in~\cite{Paulos:2011ie}.

For the five-point amplitudes, we have drastically simplified the results from~\cite{Kharel:2013mka} and expressed them in a more succinct form, as illustrated in \eqref{M5channel1} and \eqref{M5channel2}. Subsequently, we derive a recursion formula, shown in \eqref{nM}, for $n$-point amplitudes. We then apply this formula to construct higher-point calculations. More specifically, we present six-point amplitudes (refer to \eqref{M6channel2} and \eqref{M6channel3}), a seven-point amplitude (refer to \eqref{M7channel1}), and an eight-point amplitude (refer to \eqref{M8channel1}).

Besides presenting explicit novel computations, we compare our results with their flat-space counterparts. This comparison uncovers a remarkable resemblance between Mellin amplitudes and flat-space amplitudes, as detailed in the dictionary presented in Table \ref{dict}. In this table, the summation over the Mellin variables is defined in \eqref{propagator}. Additionally, the definitions of the vertex factors $\V_3$ and $\V_4$ are provided in \eqref{V33} and \eqref{V4mnnm}, respectively. It is important to note that on the Mellin side of the dictionary, an additional factor of $\frac{\pi^h}{2}\Di{n}$ should be applied.

\renewcommand{\arraystretch}{1.3}
\begin{table}
\centering
\caption{The correspondence between gluon flat-space amplitudes and Mellin amplitudes}
\label{dict}
\begin{tabular}{lcc}
\hline
{Description} & {Minkowski momentum Space} & {AdS Mellin Space} \\
\hline
Kinematic variable &$ik_i$ & $2P_i$ \\
Internal propagator & $\frac{i}{2\sum k_i\cdot k_j}$ & $\frac{1}{\tilde{\sum}\gamma_{ij}}$ \\
Three-vertex coupling& $g$ & $ g\V_3^{n_a,n_b,n_c}$ \\
Four-vertex coupling & $g^2$ & $g^2\V_4^{n_a,\cdots, n_d}$ \\
\hline
\end{tabular}
\end{table}

\section{Setting the stage: the three, four, and five-point gluon amplitudes}
\label{Section3}
We are now poised to calculate gluon amplitudes in AdS. The non-abelian gauge theory in Anti-de Sitter space is characterized by the action:
\be
S_{\texttt{YM}} = -\int d^{d+1}x \sqrt{-g} \, \frac{1}{4} \, \text{Tr}\left( F_{AB} F^{AB}\right)\ ,
\ee
where $F_{AB}^a = \partial_A A_B^a - \partial_B A_A^a +  g f^{abc} A_A^b A_B^c$ and $f^{abc}$ represent the structure constants of the gauge group. Gluon amplitudes correspond to current correlation functions and have scaling dimensions $\Delta_i=d-1$.  

\subsection{{Three-point gluon amplitude}}
The three-point gluon Mellin amplitude shown in Figure \ref{3pt}, is~\cite{Paulos:2011ie}\footnote{A detailed calculation can be found in Appendix \ref{3ptappendix}}
\begin{equation}\label{eq:3J1}
\mathscr{M}^{M_1M_2M_3}_{\texttt{3v}}= ig~\frac{ \pi^h}{2}  f^{a_1a_2a_3} \Gamma\left(\frac{\Delta_1+\Delta_2+\Delta_3-d+1}{2}\right) \left(\Di{3} \right)
\mathscr{I}_{A_1A_2A_3}\ ,
\end{equation}
where we remind the readers again that $ \Di{3} =\prod_{i=1}^3 \frac{C_{\Delta_i}}{\Gamma(\Delta_i)} \widehat{D}^{M_iA_i}$ and
\begin{equation}\label{eq:I123}
\begin{split}
\mathscr{I}_{A_1A_2A_3}&= 2\eta_{A_1A_2}\left(P_1-P_2\right)_{A_3}+\text{cyclic permutations}\ .
\end{split}
\end{equation}
Throughout this paper, we establish a  beautiful correspondence between flat-space and AdS amplitudes. As evident, already from the simple three-point function, the Mellin amplitude remarkably resembles its flat-space counterpart, obtained from the Feynman rules,
\begin{equation}
\begin{split}
\mathscr{A}_{\texttt{3v},A_1A_2A_3}&=-g f^{a_1a_2a_3} \eta_{A_1A_2}(k_1-k_2)_{A_3}+\text{cyclic permutations}\ .
\end{split}
\end{equation}
This similarity becomes apparent when the momenta $ik_{i,A}$ are mapped to $2P_{i,A}$ and the three-vertex coupling constant $g$ is associated with $g\ \V_3^{0,0,0}$, where
\be
\label{V3000}
\V_3^{0,0,0}\equiv \Gamma (d-1)\ .
\ee
The three null arguments in (\ref{V3000}) indicate that the three-vertex is linked to three boundary-to-bulk propagators. This correspondence holds, taking into account the differential operators $\Di{3}$ and a constant factor $ \frac{\pi^h}{2}$.

\begin{figure}
	\centering
	\subfigure[]{
	\begin{minipage}[t]{0.3\linewidth}
	\centering
	\includegraphics[width=1.5in]{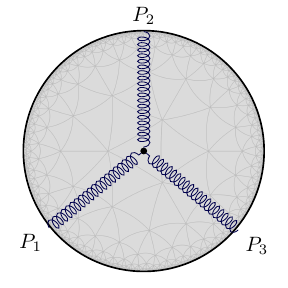}\label{3pt}
	\end{minipage}}
	\subfigure[]{
	\begin{minipage}[t]{0.3\linewidth}
	\centering
	\includegraphics[width=1.5in]{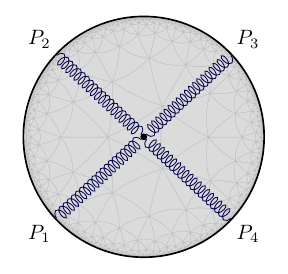}\label{4ptc}
	\end{minipage}}
	\subfigure[]{
	\begin{minipage}[t]{0.3\linewidth}
	\centering
	\includegraphics[width=1.5in]{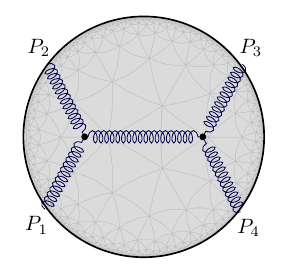}\label{4pts}
	\end{minipage}}
	\centering
\caption{\label{3 4pt} From left to right, (a) the three gluon amplitude, (b) the contact diagram of the four gluon amplitude, and (c) the $s$-channel representation of the four gluon amplitude.}
\end{figure}

 \subsection{{Four-point gluon amplitudes}}
 \subsubsection{{Contact diagram}}
\label{eq:4pt-contact}
For the four-point contact diagram (Figure {\ref{4ptc}}), the Mellin amplitude is~\cite{Kharel:2013mka}
\begin{eBox}
\begin {multline}
    \label{eq:4Jcontact1}
    \mathscr{M}^{M_1M_2M_3M_4}_{\texttt{Contact}} =-ig^2 \frac{ \pi^h}{2}  \left(f^{a_1a_4b'}f^{a_2a_3b'}+f^{a_1a_3b'}f^{a_2a_4b'}\right)\left(\Di{4}\right)\\
    \times \Gamma\left(\frac{\sum_{i=1}^4\Delta_i-d}{2}\right)\eta_{A_1A_2}\eta_{A_3A_4} +\text{cyclic perm. of (123)}\ .
\end {multline}
\end{eBox}
Note that it is also the same as the flat-space counterpart from the Feynman rules in Yang-Mills theory,
\be
\mathscr{A}_{\texttt{Contact}}=-ig^2\left(f^{a_1a_4b'}f^{a_2a_3b'}+f^{a_1a_3b'}f^{a_2a_4b'}\right)\eta_{A_1A_2}\eta_{A_3A_4} +\text{cyclic perm. of (123)}\ ,
\ee
up to $\frac{\pi^h}{2}\Di{4}$, with the identification of the four-vertex coupling constant, i.e.,
 $g^2 \V_4^{0,0,0,0}$, where
\be
\label{V40000}
\V_4^{0,0,0,0}\equiv \Gamma\left(\frac{3d-4}{2}\right)\ .
\ee
 \subsubsection{{Exchange diagram}}
 The $s$-channel is shown in Figure {\ref{4pts}}. The amplitude can be expressed in terms of the three-point function by utilizing the factorization in (\ref{eq:bbv}),
\begin {multline}
\label{exchfac}
 \Big\langle \prod_{i=1}^4J^{M_i}(P_i)\Big\rangle_{\texttt{Exch}}=\int_{-i\infty}^{i\infty}\frac{dc}{2\pi i}f_\Delta(c)\int_{\partial\text{AdS}} dQ\ \big\langle J^{M_1}(P_1)J^{M_2}(P_2)J_{h+c}^M(Q)\big\rangle\eta_{MN} \\
 \times\big\langle J_{h-c}^N(Q)J^{M_3}(P_3)J^{M_4}(P_4)\big \rangle\ ,
\end {multline}
 where the subscripts $h\pm c$ of the exchange vector field indicate its scaling dimension. The integration over $Q$ can be performed by employing Symanzik's formula~\cite{Symanzik:1972wj},
\be
\label{symanzik}
\int_{\partial\text{AdS}}  dQ\prod_{i=1}^n\Gamma(l_i)\left(-2P_i\cdot Q\right)^{-l_i}=\pi^h\int\left(\prod_{i<j}^n\frac{d\gamma_{ij}}{2\pi i}\Gamma(\gamma_{ij})P_{ij}^{-\gamma_{ij}}\right)\prod_{i=1}^n\delta\left(\sum_{j\neq i}^n\gamma_{ij}-l_i\right)
\ee
(note, $\sum_{i=1}^nl_i=d$). For review, we refer the reader to  Appendix \ref{Symanzik}.

For $Q$'s with free indices in (\ref{exchfac}), we replace all the occurrences by $P_i$'s employing (\ref{eq:eigD})~\cite{Paulos:2011ie,Kharel:2013mka}. Therefore, we get
\footnotesize
\begin{multline}
\label{eq:3J}
      \langle  J^{M_1}(P_1)  J^{M_2}(P_2) J^M_{h\pm c}(Q)\rangle
    = ig\frac{\pi^h}{2}f^{a_1a_2b} \left (\Di{2}\right)\frac{C_{h\pm c}}{\Gamma(h\pm c)}\Gamma\left(\frac{\Delta_1+\Delta_2+h\pm c-d+1}{2}\right) \\
     \times \Gamma\left(\frac{\Delta_1+\Delta_2-(h\pm c)+1}{2}\right)\Gamma\left(\frac{\Delta_1-\Delta_2+(h\pm c)+1}{2}\right)\Gamma\left(\frac{-\Delta_1+\Delta_2+h\pm c-1}{2}\right)\frac{h\pm c-1}{h\pm c}\\
     \times \Big\{X_{12}^M\Big\}\times P_{12}^{-\frac{\Delta_1+\Delta_2-(h\pm c)+1}{2}}(-2P_1\cdot Q)^{-\frac{\Delta_1-\Delta_2+h\pm c+1}{2}}(-2P_2\cdot Q)^{-\frac{-\Delta_1+\Delta_2+h\pm c-1}{2}}+(1\leftrightarrow 2)\ ,
\end{multline}
\normalsize 
where
\begin{equation}
\label{Xij}
\Big\{X_{ij}^{M}\Big\}\equiv 2\left(\eta_{A_iA_j}P_i^M-2\delta^M_{A_i}P_{i,A_j}\right)-(i\leftrightarrow j)\ .
\end{equation}
Note that in (\ref{eq:3J}) we have explicitly performed the action of the differential operator $\widehat{D}^{MA}_{h\pm c}$.

Let's take a moment to scrutinize the factor $\frac{h\pm c-1}{h\pm c}$ in (\ref{eq:3J}). One can see that it possesses simple zeros at $c=\pm (h+1)$. Importantly, with $\Delta=d-1$, the simple zeros are at the same position as the double poles of $f_\Delta(c)$. Therefore, the poles reduce to simple poles.

Integrating around one of the simple poles, say $c=h+1$ without loss of generality, one get the Mellin amplitude for the $s$-channel~\cite{Paulos:2011ie}
\begin{eBox}
\begin{multline}
\label{eq:M_s}
\mathscr{M}^{M_1M_2M_3M_4}_{\texttt{Exch}} =- g^2~\frac{\pi^h}{2} f^{a_1a_2b}f^{a_3a_4b} \left(\Di{4} \right)\\ \times \left(\sum_{n=0}^{\infty} \frac{\Big\{X_{12}^M\Big\} \V_3^{n,0,0}\times \V_3^{n,0,0}\Big\{X_{34,M}\Big\} }{4n! ~\Gamma\left(\frac{d}{2}+n\right)\left(\gamma_{12}-\frac{d}{2}+n\right)}\right)  ,
\end{multline}
\end{eBox}
where $(a)_n\equiv a(a+1)(a+2)\cdots (a+n-1)$ is the Pochhammer symbol, the contribution of the boundary points is $\{X_{ij}^M\}$ and
\be
\label{V3n00}
\V_3^{n,0,0}\equiv \left(\frac{d}{2}-n\right)_n\Gamma\left(d-1\right)
\ee
stands for the contribution of the three-vertices connecting to one bulk-to-bulk propagator. Note that with $n=0$, $\V_3^{n,0,0}$ reduces to the expression defined in (\ref{V3000}).

The series of simple poles, $\gamma_{12}=\frac{d}{2}-n$, comes from the Gamma function $\Gamma(\gamma_{12})$ in (\ref{symanzik}), where $\gamma_{12}$ should be shifted in order to incorporate the power of $P_{12}$ in (\ref{eq:3J}). Interestingly, for even $d$, the infinite sum is cut-off at $n=\frac{d}{2}$. So,
\be
\sum_{n=0}^{\infty}\frac{\V_3^{n,0,0} \times \V_3^{n,0,0}}{4n!\Gamma\left(\frac{d}{2}+n\right)\left(\gamma_{12}-\frac{d}{2}+n\right)}=&\frac{\big(\Gamma\left(d-1\right)\big)^2}{4\Gamma\left(\frac{d}{2}\right)\left(\gamma_{12}-\frac{d}{2}\right)}+\frac{\left(\Gamma\left(d-1\right)\left(\frac{d}{2}-1\right)\right)^2}{4\Gamma\left(\frac{d}{2}+1\right)\left(\gamma_{12}-\frac{d}{2}+1\right)}\\
&+\cdots \cdots \cdots+\frac{\Gamma\left(d-1\right)\left(\left(\frac{d}{2}-1\right)!\right)^2}{4(\gamma_{12}-1)}\ .
\ee

Similar to the three-point and contact examples, one can map the Mellin amplitude (\ref{eq:M_s}) from the flat-space amplitude, which from the Feynman rules is
\begin{multline}
\mathscr{A}_{\texttt{Exch}}= g^2 f^{a_1a_2b'}f^{a_3a_4b'}\frac{i}{(k_1+k_2)^2} \left(\eta_{A_1A_2}k_{1,N}-2\eta_{A_1N}k_{1,A_2}-(1\leftrightarrow 2)\right)\\
\times\left(\eta_{A_3A_4}k_{3}^N-2\eta_{A_3}^Nk_{3,A_4}-(3\leftrightarrow 4)\right)\ ,
\end{multline}
by replacing the momenta $ik_i\to 2P_i$, the propagator
\be
\frac{i}{(k_1+k_2)^2}\to \frac{1}{4n!\Gamma\left(\frac{d}{2}+n\right)\left(\gamma_{12}-\frac{d}{2}+n\right)}
\ee
(with integer $n$ to be summed from 0 to the infinity) and the three-vertex coupling constant $g\to g\  \V_3^{n,0,0}$. 

It is worth mentioning that there does not exist a unique way to express Mellin amplitudes, since it is a part of the integrand in the full correlation function (\ref{vecMellin}). For example, the term $P_1\cdot P_3$ can be absorbed into the Mellin basis with a shift of Mellin variable $\gamma_{13}\to\gamma_{13}+1$, as in~\cite{Paulos:2011ie}. However, we express the Mellin amplitude in a transparent way to show resemblance to the flat-space amplitude.

In this work, we aim to investigate Mellin gluon amplitude beyond four-point functions for different topologies. As an initial demonstration, we focus our attention on the five-point amplitudes.
\subsection{Five-point gluon amplitudes}
In this section, we explore five-point gluon AdS amplitudes in Yang-Mills theory, denoted as 
$\left\langle J^{M_1}(P_1) \cdots  J^{M_5}(P_5)\right\rangle$. These amplitudes encompass channels with two distinct types of topologies. The calculation for each five-point function involves three major steps: 1) Factorizing the five-point correlation function into a three-point function and a four-point function, implementing the integration over $Q$ using Symanzik's formula.
    2) Simplifying the double poles in the integration over $c$.
   3) Performing the integration over the Mellin variables in the four-point function.
These steps enable the computation of explicit expressions and facilitate the mapping between Mellin amplitudes and flat-space amplitudes.
\subsubsection{Channel with a three-vertex and a four-vertex}
\begin{figure}
	\centering
\includegraphics[scale=1.2]{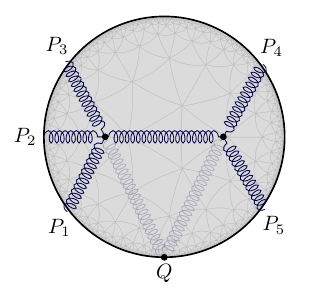}
\caption{\label{fig:5channel1}  Five-point channel consisting of contact and three-point vertex interactions}
\end{figure}
\begin{subequations}
\label{eq:equations}
We initiate our investigation by focusing on a topology, wherein a bulk-to-bulk propagator establishes a connection between a four-vertex with points $P_1$, $P_2$, and $P_3$, and a three-vertex involving points $P_4$ and $P_5$. This configuration is visually represented in Figure \ref{fig:5channel1}. Using the factorization property given by equation (\ref{eq:bbv}), the amplitude can be written as:
\begin{multline}
\label{eq:5Jchannel1}
\left\langle  \prod_{i=1}^5J^{M_i}(P_i) \right\rangle_{\texttt{3v4v}} = \int_{-i\infty}^{i\infty}\frac{dc}{2\pi i}f_\Delta(c) \int_{\partial\text{AdS}}   dQ \left\langle \prod_{i=1}^3{J}^{M_i}(P_i) J^M_{h+c}(Q)\right\rangle_{\texttt{Contact}} \\
\times\eta_{MN}\left\langle  J^N_{h-c}(Q) J^{M_4}(P_4)J^{M_5}(P_5)\right\rangle\ .
\end{multline}

Let us first look at the expression for contact diagram~\eqref{eq:4Jcontact1}. We will identify $P_4$ in~\eqref{eq:4Jcontact1} with $Q$ to match the factorized contact diagram as shown in Figure \ref{fig:5channel1}. By acting with the differential operator $\widehat{D}^{M_4A_4}$, we substitute \( P_4 \) with a free index into~\eqref{eq:4Jcontact1}. Utilizing Equation~\eqref{eq:eigD}, we can succinctly recasts the four-point contact diagram as follows~\cite{Kharel:2013mka}:
\footnotesize
\begin{multline}
\label{eq:4Jcontact}
    \left \langle \prod_{i=1}^{4} J^{M_i}(P_i) \right \rangle_{\texttt{Contact}}
 = -i g^2 ~\frac{\pi^h}{2}\left(f^{a_1a_4b'}f^{a_2a_3b'}+f^{a_1a_3b'}f^{a_2a_4b'}\right) \left( \Di{3}\right) \frac{C_4}{\Gamma(\Delta_4)} \\  
\times  \int \prod_{k< l}^4  \frac{d\gamma_{kl}'}{2\pi i} 
    \Gamma(\gamma_{kl}') P_{kl}^{-\gamma_{kl}'} \left(\frac{\Delta_4-1}{\Delta_4}\eta_{A_3}^{M_4}-\frac{1}{\Delta_4}\left(P_{1,A_3}\left( P_1^{M_4}+P_2^{M_4}+P_3^{M_4}\right)+(1\leftrightarrow 2)\right)\right) \\     \times \Gamma\left(\frac{\sum_{i=1}^4\Delta_i-d}{2}\right)\eta_{A_1A_2}+\text{cyclic perm. of (123)}\ .
\end{multline}
\normalsize
Starting from the constraint imposed by the Mellin variables, denoted as $\Delta_i' \equiv \sum_{j \neq i} \gamma_{ij}' $ (\emph{note that $\Delta_i'$ is not necessarily $\Delta_i$. For example, the term $P_{1A_3}P_1^{M_4}$ has scaling dimension of $-2$ for $P_1$. To compensate that, $\Delta_1'=\Delta_1+2$}), we have the opportunity to dispense with $ \gamma'_{i4} $ to arrive at the following constraint:
\begin{equation}
2\sum_{i<j}^3\gamma_{ij}'=\Delta_1'+\Delta_2'+\Delta_3'-\Delta_4'\ .
\end{equation}
\end{subequations}
We now transition to substituting  $P_4$ with the integration variable $Q$ in the existing formulation.

Next, we proceed by substituting (\ref{eq:3J}) and (\ref{eq:4Jcontact}) into (\ref{eq:5Jchannel1}). Employing the Symanzik's formula (\ref{symanzik}), we obtain the Mellin amplitude,
\footnotesize
\begin{multline}
\label{5M1}
\mathscr{M}^{M_1M_2M_3M_4 M_5}_{\texttt{3v4v}}
=g^3 \frac{\pi^h}{2} \left(f^{a_1bb'}f^{a_2a_3b'}+f^{a_1a_3b'}f^{a_2bb'}\right)f^{ba_4a_5}\int_{-i\infty}^{i\infty}\frac{dc}{2\pi i}\frac{c^2-(h-1)^2}{\left(c^2-(\Delta-h)^2\right)^2} \\  \times \left( \Di{5}\right)
\frac{1}{2\Gamma(c)\Gamma(-c)}\Gamma\left(\frac{\Delta_4+\Delta_5+h-c-d+1}{2}\right)\Gamma\left(\frac{\Delta_4+\Delta_5-h+c+1}{2}\right) \\ \times \ \frac{\Gamma\left(\gamma_{45}-\frac{\Delta_4+\Delta_5-h+c+1}{2}\right)}{\Gamma(\gamma_{45})}\Gamma\left(\frac{\Delta_1+\Delta_2+\Delta_3+h+c-d}{2}\right)\\ \times
\int\prod_{k<l}^3\frac{d\gamma_{kl}'}{2\pi i}\delta\left(\sum_{k<l}^3\gamma_{kl}'+\frac{h+c-\Delta'_1-\Delta'_2-\Delta'_3}{2}\right)\prod_{i<j}^3\frac{\Gamma\left(\gamma_{ij}-\gamma_{ij}'\right)\Gamma(\gamma_{ij}')}{\Gamma(\gamma_{ij})}\eta_{A_1A_2}\\ \times \Big\{X_{45,M}\Big\}\times
\left\{\eta_{A_3}^M-\frac{2}{h+c-1}\left(P_{1,A_3}\left(P_1^M+P_2^M+P_3^M\right)+(1\leftrightarrow 2)\right)\right\}+\text{cyclic perm. of (123)}\ .
\end{multline}
\normalsize
Note again that $\Delta_i'$ depends on the scaling dimension of each term in the last line. In particular, for a term which has scaling dimension of $\delta_i$ in $P_i$, $\Delta_i'=\Delta_i-\delta_i$.

Indeed, the expression \eqref{5M1} is very complicated and our goal is to simplify it further. First, we bring the external fields on shell, i.e. $\Delta_i=d-1$. The product of the $P_{4,M}$ term in $\{X_{45,M}\}$ and the second term in the second curly bracket of (\ref{5M1}) gives
\be
\label{sym45}
\eta_{A_4A_5}P_{4,M}\left(P_{1,A_3}\left(P_1^M+P_2^M+P_3^M\right)+(1\leftrightarrow 2)\right)\to\eta_{A_4A_5}P_{1,A_3}(d-1-\gamma_{45})+(1\leftrightarrow 2)\ ,
\ee
where we have shifted the Mellin variables (e.g. $\Gamma(\gamma_{14})P_{14}^{-\gamma_{14}+1}\to \gamma_{14}\Gamma(\gamma_{14})P_{14}^{-\gamma_{14}}$) and used that $\sum_{i=1}^3\gamma_{i4}+\gamma_{45}=\Delta_4=d-1$. Since (\ref{sym45}) is symmetric under the interchange of labels $4$ and $5$, its contribution vanishes due to the antisymmetric property of $\{X_{45,M}\}$.

The contribution from the multiplication of the $P_{4,A_5}$ term in $\{X_{45,M}\}$ and the second term inside the second curly bracket of (\ref{5M1}) also vanishes due to the antisymmetry under $4\leftrightarrow 5$. To see this, we can use (\ref{eigD2}) with
\begin{subequations}
\be
f(\gamma_{45})=\frac{\Gamma\left(\gamma_{45}-\frac{\Delta_4+\Delta_5-h+c+1}{2}\right)}{\Gamma(\gamma_{45})}\ ,
\ee
and get
\be
\label{part2}
f(\gamma_{45})\left(\sum_{i=1}^3P_{i,A_4}\right)P_{4,A_5}= f(\gamma_{45})\frac{1}{2}\eta_{A_4A_5}-f(\gamma_{45}-1)P_{4,A_5}P_{5,A_4}+\cdots\ ,
\ee
\end{subequations}
where $\cdots$ denotes the term vanishing upon the action of $\widehat{D}^{M_4A_4}$. Now it is clear that (\ref{part2}) exhibits symmetry under $4\leftrightarrow 5$. Therefore, the antisymmetric property of $\{X_{45,M}\}$ results in a total cancellation. As a result of the above observations, we find that the second term inside the second curly bracket of (\ref{5M1}) can be neglected. 

We also note that with $\Delta=d-1$, the factor
\be
\frac{c^2-(h-1)^2}{(\left(c^2-(\Delta-h)^2\right)^2}=\frac{1}{c^2-(h-1)^2}\ ,
\ee
which gives simple poles at $c=\pm(h-1)$. Therefore, with the integration of $c$ around the pole $h-1$, \eqref{5M1} can be simplified to:
\footnotesize
\begin{multline}
\label{5M1sim}
\mathscr{M}^{M_1M_2M_3M_4 M_5}_{\texttt{3v4v}}
=g^3 \frac{\pi^h}{2} \left(f^{a_1bb'}f^{a_2a_3b'}+f^{a_1a_3b'}f^{a_2bb'}\right)f^{ba_4a_5}\left( \Di{5}\right)
\frac{\Gamma\left(d-1\right)}{4}\\ \times
\frac{\Gamma\left(\gamma_{45}-d+1\right)}{\Gamma(\gamma_{45})\Gamma\left(1-\frac{d}{2}\right)}\int\left(\prod_{k<l}^3\frac{d\gamma_{kl}'}{2\pi i}\frac{\Gamma\left(\gamma_{kl}-\gamma_{kl}'\right)\Gamma(\gamma_{kl}')}{\Gamma(\gamma_{kl})}\right)\delta\left(\sum_{k<l}^3\gamma_{kl}'-d+1\right)\\
\times\Gamma\left(\frac{3d-4}{2}\right) \eta_{A_1A_2}\Big\{X_{45,A_3}\Big\}+\text{cyclic perm. of (123)}\ .
\end{multline}
\normalsize
After integrating (\ref{5M1sim}) over the Mellin variables $\gamma_{kl}'$, for $1\le k,l\le 3$, we encounter poles at $\gamma_{kl}'=\gamma_{kl}+n_{kl}$, with any non-negative integers $n_{kl}$. However, because of the presence of a delta function, one of these pole terms does not get integrated out.\footnote{Note that the scalar case has a similar construction~\cite{Fitzpatrick:2011ia}.}
\begin{subequations}
To compare the expression \eqref{5M1sim} to flat-space amplitude, we need to simplify it even further. We begin by considering the integral term in the second line of \eqref{5M1sim}:
\footnotesize
\begin{equation}\label{eq:start}
\int\left(\prod_{k<l}^3\frac{d\gamma_{kl}'}{2\pi i}\frac{\Gamma\left(\gamma_{kl}-\gamma_{kl}'\right)\Gamma(\gamma_{kl}')}{\Gamma(\gamma_{kl})}\right)\delta\left(\sum_{k<l}^3\gamma_{kl}'-d+1\right)
\end{equation}
\normalsize
Through the integration around the poles $\gamma_{kl}'=\gamma_{kl}+n_{kl}$, this term reduces to:
\footnotesize
\begin{equation}\label{eq:mid}
\sum_{m=0}^\infty\frac{1}{\sum_{i<j}^3\gamma_{ij}-d+1+m}\sum_{\sum_{i<j}^3n_{ij}=m}(-1)^m\frac{(\gamma_{12})_{n_{12}}(\gamma_{13})_{n_{13}}(\gamma_{23})_{n_{23}}}{n_{12}!n_{13}!n_{23}!}
\end{equation}
\normalsize
Further simplification leads it to:
\footnotesize
\begin{equation}\label{eq:final1}
\sum_{m=0}^\infty\frac{(-1)^m}{\sum_{i<j}^3\gamma_{ij}-d+1+m}\frac{\left(\sum_{i<j}^3\gamma_{ij}\right)_m}{m!}= \sum_{m=0}^\infty\frac{(-1)^m\left(d-1-m\right)_m}{m!(\sum_{i<j}^3\gamma_{ij}-d+1+m)}
\end{equation}
\normalsize
At the pole of \eqref{eq:final1}, $\sum_{i<j}^3\gamma_{ij}=d-1-m$, from the constraint $\sum_{j\neq i}\delta_{ij}=\Delta_i'$ on the Mellin variables and $\Delta_i=d-1$, we have
\footnotesize
\begin{equation}
\label{pole}
\gamma_{45}=\frac{\Delta_4+\Delta_5+1-\sum_{i=1}^3\Delta_i+2\sum_{i<j}^3\gamma_{ij}}{2}=\frac{d}{2}-m\ .
\end{equation}
\normalsize
Hence, the first term in the second line of \eqref{5M1sim} becomes
\footnotesize
\begin{equation}
\begin{split}
\frac{\Gamma\left(\gamma_{45}-d+1\right)}{\Gamma\left(\gamma_{45}\right)\Gamma\left(1-\frac{d}{2}\right)}=\frac{(-1)^m\left(\frac{d}{2}-m\right)_m}{\Gamma\left(\frac{d}{2}+m\right)}\ .
\end{split}
\end{equation}
\normalsize
\end{subequations}

Finally, the Mellin amplitude becomes remarkably simple:
\begin{eBox}
\begin{multline}
\label{M5channel1}
\mathscr{M}^{M_1\cdots M_5}_{\texttt{3v4v}}
=g^3 \frac{\pi^h}{2} \left(f^{a_1bb'}f^{a_2a_3b'}+f^{a_1a_3b'}f^{a_2bb'}\right)f^{ba_4a_5} \left( \Di{5}\right)\\
\times \sum_{m=0}^\infty\frac{\Big\{X_{45,A_3} \Big\}\V_3^{m,0,0} \times \V_4^{m,0,0,0}}{4m!\Gamma\left(\frac{d}{2}+m\right)(\gamma_{45}-\frac{d}{2}+m)}\eta_{A_1A_2}+\text{cyclic perm. of (123)}\ ,
\end{multline}
\end{eBox}
where we have defined
\be
\label{V4m000}
\V_4^{m,0,0,0}\equiv (d-1-m)_m\Gamma\left(\frac{3d-4}{2}\right)\ .
\ee
Note that with $m=0$, it reduces to $\V_4^{0,0,0,0}$ given in (\ref{V40000}). We should remark that the summation over $m$ is truncated at  $m = d-1$ for odd values of $d$, and at $m = \min\{d-1, \frac{d}{2}\}$ for even $d$. With this stipulation in place, we are now in an ideal position to compare~\eqref{M5channel1} with its corresponding flat-space expression. From the Feynman rules,
\begin{multline}
\mathscr{A}_{\texttt{3v4v}}=ig^3\left(f^{a_1bb'}f^{a_2a_3b'}+f^{a_1a_3b'}f^{a_2bb'}\right)f^{ba_4a_5}\left(\eta_{A_4A_5}k_{4,A_3}-2\eta_{A_3A_4}k_{4A_5}-(4\leftrightarrow 5)\right) \\ \times\eta_{A_1A_2}\frac{i}{(k_4+k_5)^2}
+\text{cyclic perm. of (123)}\ .
\end{multline}
One can immediately see that, the simplification seen at three and four-point gluon amplitudes, as elaborated in Section~\ref{Section3} carries for the higher-point structure with three-vertex and four-vertex topology. Specifically, we find that the Mellin amplitude in the current channel can be straightforwardly derived from its flat-space counterpart through a well-defined set of substitutions. Explicitly, for any momentum term \( ik_{i} \), it should be replaced by \( 2P_{i} \).
For the propagator,
\begin{equation}
\frac{i}{(k_4+k_5)^2}\to \frac{1}{4\left(\gamma_{45}-\frac{d}{2}+m\right)\Gamma\left(\frac{d}{2}+m\right)m!}\ ,
\end{equation}
with $m$ to be summed over. 
For the three- and four-vertex coupling constant,
\be
g \mapsto g\ \V_3^{m,0,0}\ \quad \text{and} \quad g^2 \mapsto g^2\ \V_4^{m,0,0,0} \ ,
\ee
respectively. In summary, our exhaustive computational analysis reveals a striking simplification in the Mellin amplitude associated with the five-point function. In the ensuing section, we will delve into the other topological configurations that constitute this five-point function.
\subsubsection{Channel with three three-vertices}
\begin{figure}
	\centering
\includegraphics[scale=1.2]{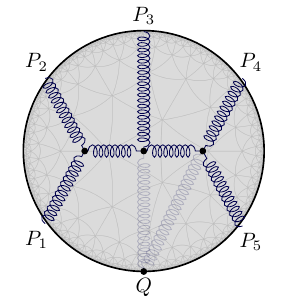}
\caption{\label{5channel2} The channel of five-point gluon amplitude for (\ref{5Jchannel2})}
\end{figure}
In this part of the paper, we focus on the other five-point channel configuration where there are three-vertices, one links $P_1$, $P_2$ with a bulk-to-bulk propagator, one links $P_4$, $P_5$ with another bulk-to-bulk propagator, and the other links $P_3$ with the two bulk-to-bulk propagators. This diagram depicted in Figure \ref{5channel2}. By employing the relationship specified in (\ref{eq:bbv}), we can derive the corresponding amplitude for this configuration,
\begin{subequations}
\begin{multline}
\label{5Jchannel2}
\left\langle  \prod_{i=1}^5J^{M_i}(P_i)\right\rangle_{\texttt{3v3v3v}}= \int_{-i\infty}^{i\infty}\frac{dc}{2\pi i}f_\Delta(c) \int_{\partial\text{AdS}} dQ \left\langle \prod_{i=1}^3 J^{M_i}(P_i) J^M_{h+c}(Q)\right\rangle_{\texttt{Exch}}\\
\times\eta_{MN}\left\langle  J^N_{h-c}(Q) J^{M_4}(P_4)J^{M_5}(P_5)\right\rangle\ .
\end{multline}
The exchange contribution, with $\Delta_i=d-1$, is given by~\eqref{eq:M_s}. Performing the action of $\widehat{D}^{M_4A_4}$ explicitly, one rewrite the expression as
\footnotesize
\begin{multline}
\label{4Js}
\bigg\langle \prod_{i=1}^4 J^{M_i}(P_i)\bigg\rangle_{{\texttt{Exch}}}
=-g^2f^{a_1a_2b'}f^{a_3a_4b'}\frac{\pi^h}{2}  \left(\Di{3} \right) \frac{C_4}{\Gamma(\Delta_4)}\int\prod_{k<l}^4\frac{d\gamma_{kl}'}{2\pi i}\Gamma (\gamma_{kl}')P_{kl}^{-\gamma_{kl}'}\\
\times \left(\sum_{n=0}^{\infty}\frac{\left(\Gamma\left(d-1\right)(\frac{d}{2}-n)_n\right)^2}{4n!\ \Gamma\left(\frac{d}{2}+n\right)\left(\gamma'_{12}-\frac{d}{2}+n\right)}\right) \left(\frac{\Delta_4-1}{\Delta_4}\eta^{M_4A_b} \mathscr{G}_{A_1A_2A_3A_b}+\frac{2}{\Delta_4}\bigg(P_1^{M_4}+P_2^{M_4}+P_3^{M_4}\bigg) \mathscr{H}_{A_1A_2A_3}\right)\ ,
\end{multline}
\normalsize
where
\begin{equation}
\label{G}
\mathscr{G}_{A_1A_2A_3A_b}
=\Big\{X_{12}^{A_{b'}}\Big\}\Big\{X_{bb',A_3}\Big\}\ ,\ P_b\equiv-P_1-P_2-P_3\ , \ P_{b'}\equiv P_1+P_2\ ,
\end{equation}
and $\mathscr{H}_{A_1A_2A_3}$ is obtained from the product $P_4^{A_b} \mathscr{G}_{A_1A_2A_3A_b}$ followed by the elimination of the $P_4$ dependence using (\ref{eigD1}) and (\ref{eigD2}).

\end{subequations}
After substituting (\ref{eq:3J}) and (\ref{4Js}) into (\ref{5Jchannel2}), we proceed to integrate over $Q$ by using the Symanzik's formula (\ref{symanzik}). Upon completing the integration, the resulting expression yields the Mellin amplitude, 
\footnotesize
\begin{multline}
\label{5M2}
\mathscr{M}^{M_1\cdots M_5}_{\texttt{3v3v3v}}
=-ig^3f^{a_1a_2b'}f^{a_3bb'}f^{ba_4a_5}\frac{\pi^h}{2}\int_{-i\infty}^{i\infty}\frac{dc}{2\pi i}\frac{c^2-(h-1)^2}{\left(c^2-(\Delta-h)^2\right)^2} \left( \Di{5} \right) 
\frac{1}{2\Gamma(c)\Gamma(-c)} \\ \times \Gamma\left(\frac{\Delta_4+\Delta_5+h-c-d+1}{2}\right)\Gamma\left(\frac{\Delta_4+\Delta_5-h+c+1}{2}\right)
\frac{\Gamma\left(\gamma_{45}-\frac{\Delta_4+\Delta_5-h+c+1}{2}\right)}{\Gamma(\gamma_{45})}\\
\times\int\left(\prod_{k<l}^3\frac{d\gamma_{kl}'}{2\pi i}\frac{\Gamma\left(\gamma_{kl}-\gamma_{kl}'\right)\Gamma(\gamma_{kl}')}{\Gamma(\gamma_{kl})} \right)\delta\left(\sum_{ k<l}^3\gamma_{kl}'+\frac{h+c-\sum_{i=1}^3\Delta_i'}{2}\right)
\\ \times \sum_{n=0}^{\infty}\frac{\left(\Gamma\left(d-1\right)(-\frac{d}{2}+1)_n\right)^2}{4n!\Gamma\left(\frac{d}{2}+n\right)\left(\gamma_{12}'-\frac{d}{2}+n\right)}\Big\{X_{45}^M\Big\}
\Big\{\mathscr{G}_{A_1A_2A_3M}+\frac{2}{h+c-1} (P_1+P_2+P_3)_M\mathscr{H}_{A_1A_2A_3}\Big\}\ .
\end{multline}
\normalsize
Note again that here $\Delta_i'$ is not necessarily $\Delta_i$. From the expression of $\mathscr{G}_{A_1A_2A_3M}$, (\ref{G}), $\sum_{i=1}^3\Delta_i'=\sum_{i=1}^3\Delta_i+2$ for the $\mathscr{G}$ term, and $\sum_{i=1}^3\Delta_i'=\sum_{i=1}^3\Delta_i+4$ for the $\mathscr{H}$ term.

Now we simplify the expression \eqref{5M2} further. Taking $\Delta_i=d-1$, the result obtained by multiplying $\eta_{A_4A_5}P_4^M$ term from $\{X_{45}^M\}$ with the second term inside the second curly bracket of (\ref{5M2}) is (similar to (\ref{sym45})), :
\be
\label{part12}
\frac{2}{h+c-1}\eta_{A_4A_5}(d-1-\gamma_{45})\mathscr{H}_{A_1A_2A_3}(P_1,P_2,P_3)\ ,
\ee
where we have used that $\sum_{i=1}^3\gamma_{i4}+\gamma_{45}=\Delta_4=d-1$. Since (\ref{part12}) is symmetric under $4\leftrightarrow 5$, its contribution vanishes due to the antisymmetric property of $\{X_{45,M}\}$. The product of $\eta_{A_4M}P_{4,A_5}$ in $\{X_{45}^M\}$, the second term inside the second curly bracket of (\ref{5M2}) and 
\begin{subequations}
\be
f(\gamma_{45})=\frac{\Gamma\left(\gamma_{45}-\frac{\Delta_4+\Delta_5-h+c+1}{2}\right)}{\Gamma(\gamma_{45})}\ ,
\ee
similar to (\ref{part2}), gives
\begin{multline}
\label{part22}
f(\gamma_{45})\left(P_{1,A_4}+P_{2,A_4}+P_{3,A_4}\right)P_{4,A_5}\mathscr{H}_{A_1A_2A_3}(P_1,P_2,P_3)= \\
\left(f(\gamma_{45})\frac{1}{2}\eta_{A_4A_5}-f(\gamma_{45}-1)P_{4,A_5}P_{5,A_4}\right)\mathscr{H}_{A_1A_2A_3}(P_1,P_2,P_3)+\cdots\ ,
\end{multline}
\end{subequations}
where $\cdots$ denotes the term vanishing upon the action of $\widehat{D}^{M_4A_4}$. The expression (\ref{part22}) exhibits symmetry under $4\leftrightarrow 5$. As a result, again due to the antisymmetry of $\{X_{45}^M\}$, (\ref{part22}) gets canceled out. Consequently, based on the preceding arguments, we can conclude that the $\mathscr{H}_{A_1A_2A_3}$ term of (\ref{5M2}) can be safely disregarded. Then, the same as in the other channel, the poles of $c$ become simple. And we can integrate around the pole $c=h-1$.

Besides, the pole of $\gamma_{ij}'$ (for $1\le i<j\le 3$) is at $\gamma_{ij}+n_{ij}$, with any non-negative integers $n_{ij}$. Integrating over it around the pole, we have
\begin{subequations}
\footnotesize
\begin{multline}
\label{sim}
\int\prod_{k<l}^3\frac{d\gamma_{kl}'}{2\pi i}\ \delta\left(\sum_{k<l}^3\gamma_{kl}'-d\right)\prod_{i<j}^3\frac{\Gamma\left(\gamma_{ij}-\gamma_{ij}'\right)\Gamma(\gamma_{ij}')}{\Gamma(\gamma_{ij})}\frac{1}{\gamma_{12}'-\frac{d}{2}+n}\\
=\sum_{m=0}^\infty\frac{1}{\sum_{i<j}^3\gamma_{ij}-d+m}\sum_{\sum_{i<j}^3n_{ij}=m}(-1)^m\frac{(\gamma_{12})_{n_{12}}(\gamma_{13})_{n_{13}}(\gamma_{23})_{n_{23}}}{n_{12}!n_{13}!n_{23}!}\frac{1}{\gamma_{12}+n_{12}-\frac{d}{2}+n}\\
=\sum_{m=0}^\infty\frac{1}{\sum_{i<j}^3\gamma_{ij}-d+m}\sum_{n_{12}=0}^m(-1)^m\frac{(\gamma_{12})_{n_{12}}(\gamma_{13}+\gamma_{23})_{m-n_{12}}}{n_{12}!(m-n_{12})!}\frac{1}{\gamma_{12}+n_{12}-\frac{d}{2}+n}\ .
\end{multline}
\normalsize
The above equation has poles at
\begin{equation}
\label{2pole}
\sum_{i<j}^3 \gamma_{ij} = d - m \quad \text{and} \quad \gamma_{12} = \frac{d}{2} - n - n_{12}.
\end{equation}
Using (\ref{2pole}) and shifting $n\to n-n_{12}$, we get
\begin{multline}
\sum_{n_{12}=0}^m\frac{(\gamma_{12})_{n_{12}}(\gamma_{13}+\gamma_{23})_{m-n_{12}}}{n_{12}!(m-n_{12})!}\frac{\left(\Gamma(d-1)(\frac{d}{2}-n)_n\right)^2}{4n!\Gamma\left(\frac{d}{2}+n\right)\left(\gamma_{12}+n_{12}-\frac{d}{2}+n\right)}\\
\to \V_3^{m,n,0}\frac{\Gamma(d-1)(\frac{d}{2}-n)_n}{4m!n!\Gamma\left(\frac{d}{2}+n\right)\left(\gamma_{12}-\frac{d}{2}+n\right)}\ ,
\end{multline}
where
\be
\label{V3mn0}
\V_3^{m,n,0}\equiv &\Gamma(d-1)m!\sum_{n_{12}=0}^{\min\{m,n\}}\frac{(\frac{d}{2}-m+n)_m(\frac{d}{2}-n+n_{12})_{n-n_{12}}(n-n_{12}+1)_{n_{12}}}{n_{12}!(m-n_{12})!}\\
=&\Gamma(d-1)\left(\frac{d}{2}-n+m\right)_n\left(\frac{d}{2}-m+n\right)_m\ .
\ee
\end{subequations}
Note that $\V_3^{m,n,0}$ is explicitly symmetric under $m\leftrightarrow n$. It can also be easily checked that with $n=0$, $\V_3^{m,0,0}$ reduces to (\ref{V3n00}).  Now, we can rewrite the Mellin amplitude in the simplified form
\begin{eBox}
\begin{multline}
\label{M5channel2}
\mathscr{M}^{M_1\cdots M_5}_{\texttt{3v3v3v}}=-ig^3f^{a_1a_2b'}f^{a_3bb'}f^{ba_4a_5}\frac{\pi^h}{2} \left(\Di{5}\right)\sum_{m,n=0}^\infty \Big\{X_{bb',A_3}\Big\}\V_3^{m,n,0}\\
\times\frac{\Big\{X_{45}^{A_b}\Big\}\V_3^{m,0,0}}{4m!\Gamma\left(\frac{d}{2}+m\right)\left(\gamma_{45}-\frac{d}{2}+m\right)}  \times \frac{\Big\{X_{12}^{A_{b'}}\Big\}\V_3^{n,0,0}}{4n!\Gamma\left(\frac{d}{2}+n\right)\left(\gamma_{12}-\frac{d}{2}+n\right)}\ .
\end{multline}
\end{eBox}
For even $d$, the infinite sums are cut-off at $m=\frac{d}{2}$ and $n=\frac{d}{2}$, and only finite number of terms remain.

Just as in the channel (\ref{M5channel1}), we can compare (\ref{M5channel2}) to its flat-space counterpart,
\begin{multline}
\mathscr{A}_{\texttt{3v3v3v}}=-g^3f^{a_1a_2b'}f^{a_3bb'}f^{ba_4a_5}\left(\eta_{MM'}q_{A_3}-2\eta_{MA_3}q_{M'}-(qM\leftrightarrow q'M')\right)\\
\times\left(\eta_{A_4A_5}k_4^M-2\eta_{A_4}^Mk_{4,A_5}-(4\leftrightarrow 5)\right)\frac{i}{(k_4+k_5)^2}\\
\times\left(\eta_{A_1A_2}k_1^{M'}-2\eta_{A_1}^{M'}k_{1,A_2}-(1\leftrightarrow 2)\right)\frac{i}{(k_1+k_2)^2}\ ,
\end{multline}
with $q\equiv -k_1-k_2-k_3$ and $q'\equiv k_1+k_2$. The dictionary between the Mellin amplitude and the flat-space amplitude, which is obtained from the Feynman rules, can be read off as follows. For the momenta, $ik_i\to 2P_i$. For the propagator
\begin{equation}
\frac{i}{(k_4+k_5)^2}\to \frac{1}{4m!\Gamma\left(\frac{d}{2}+m\right)\left(\gamma_{45}-\frac{d}{2}+m\right)}\ ,
\end{equation}
with $m$ to be summed over. And
\begin{equation}
\frac{i}{(k_1+k_2)^2}\to \frac{1}{4n!\Gamma\left(\frac{d}{2}+n\right)\left(\gamma_{12}-\frac{d}{2}+n\right)}\ ,
\end{equation}
with $n$ to be summed over. For the three-vertex coupling constant, 
\be
g\to g\ \V_3^{m,0,0},\quad g\ \V_3^{n,0,0}, \quad g\ \V_3^{m,n,0} \ .
\ee
Note that the former two, each of which has two zero arguments, are for the three-vertices connected to two external fields, while the last one, which has only one zero argument, is for the three-vertex connected to only one external fields.
\section{Higher-point gluon amplitudes}
\label{Section4}
\subsection{Factorization of $(n+1)$-point gluon amplitudes}
\begin{figure}
	\centering
\includegraphics[scale=1.2]{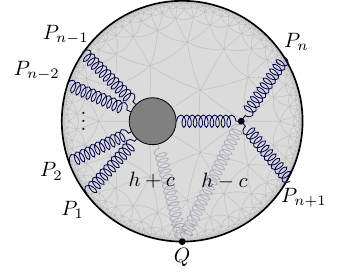}
\caption{\label{nchannel} A $(n+1)$-current amplitude for involving a three-vertex}
\end{figure}
So far we have presented lower-point gluon amplitudes (three to five-point) that already exist in the literature. Particularly, we have considerably simplified the five-point result. There we demonstrated how the calculation of the five-point gluon amplitude can be done by factorizing it into a four-point amplitude and a three-point amplitude. One interesting finding is that the antisymmetry between the two external legs $ P_4$ and $ P_5$, which are brought together to a three-vertex, kills the term $\mathscr{H}_{A_1A_2A_3}$. And the remaining term has simple poles at $c=\pm(h-1)$. 

It is important to observe that this simplification does not depend on the details of the four-point amplitude, i.e. $\mathscr{G}_{A_1A_2A_3M}$ and $\mathscr{H}_{A_1A_2A_3}$ in (\ref{4Js}). So, presumably for any $(n+1)$-point gluon amplitude factorized into an $n$-point amplitude and a three-vertex (see Figure \ref{nchannel}), we expect a similar simplification. In this section we will explicitly demonstrate this fact and apply it in higher-point computations.

First, from the factorization of bulk-to-bulk propagator (\ref{eq:bbv}), we can calculate the $(n+1)$-point gluon amplitude from the lower-point amplitudes.
\begin{multline}
\label{nJchannel}
\left \langle   \prod_{i=1}^{n+1}J^{M_i}(P_i)\right \rangle= \int_{-i\infty}^{i\infty}\frac{dc}{2\pi i}f_\Delta(c)\int_{\partial\text{AdS}} dQ \left \langle \prod_{i=1}^{n-1}  J^{M_i}(P_i)  J^M_{h+c}(Q) \right \rangle\\
\times\eta_{MN}\left\langle   J^N_{h-c}(Q)  J^{M_n}(P_n)  J^{M_{n+1}}(P_{n+1})\right\rangle\ .
\end{multline}\

An $n$-point Mellin amplitude can be written in the following form.
\begin{multline}
\label{n1J}
\mathscr{M}^{M_1M_2\cdots M_n}=\frac{\pi^h}{2}\left(\Di{n}\right) \tilde{\mathscr{M}}^{a_1a_2\cdots a_n}_{A_1A_2\cdots A_n} 
=\frac{\pi^h}{2} \left(\Di{n-1}\right)\frac{C_{n}}{\Gamma(\Delta_{n})}  \\ \left(\frac{\Delta_{n}-1}{\Delta_{n}}\eta^{M_nA_n}\tilde{\mathscr{M}}^{a_1a_2\cdots a_n}_{A_1A_2\cdots A_n} 
+\frac{1}{\Delta_n}\frac{\partial}{\partial P_n^{M_n}}\mathscr{H}^{a_1a_2\cdots a_n}_{A_1A_2\cdots A_{n-1}}\right)\ ,
\end{multline}
where $\mathscr{H}^{a_1a_2\cdots a_n}_{A_1A_2\cdots A_{n-1}}\equiv P_n^{A_n}\tilde{\mathscr{M}}^{a_1a_2\cdots a_n}_{A_1A_2\cdots A_n}$.

As in the examples (\ref{eigD1}), (\ref{eigD2}) and their generalizations with more free indices, we can always use (\ref{eq:eigD}) to replace $P_n^A$ by the other $P_i^A$'s. In this way, we can eliminate the dependences of $\tilde{\mathscr{M}}_{A_1A_2\cdots A_n}$ and $\mathscr{H}_{A_1A_2\cdots A_{n-1}}$ on $P_n$. Then, the second term in the last line of (\ref{n1J}) can be further calculated as follows.
\begin{multline}
\label{QQG}
\frac{\partial}{\partial P_n^{M_n}}\mathscr{H}_{A_1A_2\cdots A_{n-1}}(P_1,P_2,\cdots,P_{n-1})\prod_{k<l}^n\Gamma(\gamma_{kl}')P_{kl}^{-\gamma_{kl}'}
= \\ 2\sum_{i=1}^{n-1}P_i^{M_n}\mathscr{H}_{A_1A_2\cdots A_{n-1}}(\gamma_{in}'\to\gamma_{in}'-1)\prod_{k<l}^n\Gamma(\gamma_{kl}')P_{kl}^{-\gamma_{kl}'}\ ,
\end{multline}
where $\gamma_{ij}'$ denote the Mellin variables for the lower point Mellin amplitude $\mathscr{M}_n$. Plugging (\ref{eq:3J}), (\ref{n1J}) and (\ref{QQG}) in (\ref{nJchannel}), we have
\footnotesize
\begin{multline}
\label{nM1}
\mathscr{M}^{M_1\cdots M_n}
=igf^{a_na_{n+1}b}\frac{\pi^h}{2}\int_{-i\infty}^{i\infty}\frac{dc}{2\pi i}\frac{c^2-(h-1)^2}{\left(c^2-(\Delta-h)^2\right)^2}\left(\Di{n+1} \right)\frac{1}{2\Gamma(c)\Gamma(-c)}\\ \times 
\Gamma\left(\frac{\Delta_n+\Delta_{n+1}+h-c-d+1}{2}\right)\Gamma\left(\frac{\Delta_n+\Delta_{n+1}-h+c+1}{2}\right) \frac{\Gamma\left(\gamma_{n(n+1)}-\frac{\Delta_n+\Delta_{n+1}-h+c+1}{2}\right)}{\Gamma(\gamma_{n(n+1)})}\\\times\int\prod_{i<j}^{n-1}\frac{d\gamma_{ij}'}{2\pi i}\frac{\Gamma\left(\gamma_{ij}-\gamma_{ij}'\right)\Gamma(\gamma_{ij}')}{\Gamma(\gamma_{ij})}\delta\left(\sum_{ k<l}^{n-1}\gamma_{kl}'+\frac{h+c-\sum_{i=1}^{n-1}\Delta_i'}{2}\right)\\\times \Big\{X_{n(n+1)}^M\Big\}\left\{\tilde{\mathscr{M}}_{A_1A_2\cdots A_{n-1}M}^{a_1a_2\cdots a_{n-1}b}(\gamma_{ij}')+\frac{2}{h+c-1}\sum_{i=1}^{n-1}P_{i,M}\mathscr{H}^{a_1a_2\cdots a_{n-1}b}_{A_1A_2\cdots A_{n-1}}(\gamma_{ij}')\right\}\ .
\end{multline}
\normalsize
Again $\Delta_i'\equiv \Delta_i-\delta_i$ where $\delta_i$ denotes the scaling dimension of $P_i$ for each term in the last line. Note that for all the terms in the sum in the last line, they have the same total scaling dimension $\sum_i^{n-1}\Delta_i'$.

Now we take $\Delta_i=d-1$ and show that the contribution of $\mathscr{H}_{A_1A_2\cdots A_{n-1}}$ vanishes. First,
\begin{multline}
\label{part126}
\eta_{A_nA_{n+1}}P_n^M\sum_{i=1}^{n-1}P_{i,M}\mathscr{H}_{A_1A_2\cdots A_{n-1}}\to-\frac{1}{2}\eta_{A_nA_{n+1}}(d-1-\gamma_{n(n+1)})\mathscr{H}_{A_1A_2\cdots A_{n-1}}\ ,
\end{multline}
where we have shifted the Mellin variables $\gamma_{in}\to \gamma_{in}+1$ (no prime) for each term in the sum and used that $\sum_{i=1}^{n-1}\gamma_{in}+\gamma_{n(n+1)}=\Delta_n=d-1$. Since (\ref{part126}) is symmetric under the interchange of labels $n$ and $n+1$ (i.e., $n\leftrightarrow n+1$), the contribution of it vanishes due to the antisymmetry of $\{X_{n(n+1}^M\}$. Second, using (\ref{eigD2}) with
\begin{subequations}
\be
f(\gamma_{n(n+1)})=\frac{\Gamma\left(\gamma_{n(n+1)}-\frac{\Delta_n+\Delta_{n+1}-h+c+1}{2}\right)}{\Gamma(\gamma_{n(n+1)})}\ ,
\ee
we can get
\begin{multline}
\label{part226}
f(\gamma_{n(n+1)})\sum_{i=1}^{n-1}P_{i,A_n}P_{n,A_{n+1}}\mathscr{H}_{A_1A_2\cdots A_{n-1}}\\
=\left(\frac{1}{2}f(\gamma_{n(n+1)})\eta_{A_nA_{n+1}}-f(\gamma_{n(n+1)}-1)P_{n,A_{n+1}}P_{n+1,A_n}\right)\mathscr{H}_{A_1A_2\cdots A_{n-1}} +\cdots\ ,
\end{multline}
\end{subequations}
where $\cdots$ denotes the term vanishing upon the action of $\widehat{D}^{M_nA_n}$. The expression (\ref{part226}) again exhibits symmetry under $n\leftrightarrow n+1$. As a result, its contribution gets completely canceled out from the antisymmetry of $\{X_{n(n+1}^M\}$. Combining these, we can conclude that the $\mathscr{H}$ term in (\ref{nM1}) vanishes.

Finally, the integration around the simple pole $c=\Delta-h$ (with $\Delta=d-1$) and the poles $\gamma_{ij}'=\gamma_{ij}+n_{ij}$ with any non-negative integers $n_{ij}$ gives the important recursion formula from an $n$-point amplitude to an $(n+1)$-point amplitude.
\begin{eBox}
\begin{multline}
\label{nM}
\mathscr{M}^{M_1M_2\cdots M_{n+1}}_{n+1}
=\frac{\pi^h}{2} \left(\Di{n+1}\right)igf^{a_na_{n+1}b}\sum_{m=0}^\infty\frac{\Big\{X_{n(n+1)}^M\Big\}\V_3^{m,0,0}}{4\Gamma\left(\frac{d}{2}+m\right)(\gamma_{n(n+1)}-\frac{d}{2}+m)}\\ \times\sum_{\sum_{i<j}^{n-1}n_{ij}=m}\prod_{i<j}^{n-1}\frac{(\gamma_{ij})_{n_{ij}}}{n_{ij}!}
\tilde{\mathscr{M}}_{n,A_1A_2\cdots A_{n-1}M}^{a_1a_2\cdots a_{n-1}b}(P_1,P_2,\cdots,P_{n-1})\big|_{\gamma_{ij}'\to\gamma_{ij}+n_{ij}}\ .
\end{multline}
\end{eBox}
One can use this recursion formula to calculate any $n$-point gluon amplitudes with at most one four-vertex. For example, if a channel contains one four-vertex, we can start from the four-vertex and add three-vertices one by one. And at each step of adding one more three-vertex, the amplitude can be calculated by using (\ref{nM}).
\subsection{Six-point amplitude: snowflake channel}
\begin{figure}
	\centering
\includegraphics[scale=1.2]{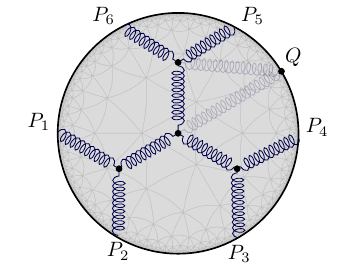}
\caption{\label{6channel2} The ``snowflake'' channel of six point gluon amplitude for (\ref{6Jchannel2})}
\end{figure}
We can use (\ref{nM}) to calculate the six-point gluon Mellin amplitude for the diagram in Figure \ref{6channel2}. As shown in the figure, the amplitude can be factorized into a five-point gluon amplitude and a three-point gluon amplitude. Therefore,
\begin{multline}
\label{6Jchannel2}
\mathscr{M}_{\texttt{Snowflake}}^{M_1M_2\cdots M_6}
=\frac{\pi^h}{2} \left(\Di{6}\right)igf^{a_5a_6b}\Big\{X_{56}^M\Big\}\sum_{m=0}^\infty\frac{\V_3^{m,0,0}}{4\Gamma\left(\frac{d}{2}+m\right)(\gamma_{56}-\frac{d}{2}+m)}\\ \times
\sum_{\sum_{i<j}^4n_{ij}=m}\prod_{i<j}^4\frac{(\gamma_{ij})_{n_{ij}}}{n_{ij}!}\tilde{\mathscr{M}}_{\texttt{3v3v3v},A_1A_2\cdots A_4M}^{a_1a_2a_3a_4b}(P_1,P_2,P_3,P_4)\big|_{\gamma_{ij}'\to\gamma_{ij}+n_{ij}}\ .
\end{multline}

In (\ref{6Jchannel2}) $\tilde{\mathscr{M}}_{\texttt{3v3v3v}}$ can achieved by eliminating the $P_3$ dependence in the Mellin amplitude (\ref{M5channel2}), interchanging the labels $1\leftrightarrow 2$ and $3\leftrightarrow 5$, and replacing $(A_5,a_5)$ by $(M,b)$. We first eliminate $P_3$ in the Mellin amplitude. One appearance of $P_3$ is either in the inner products $P_i\cdot P_3-(i\leftrightarrow j)$ (where $(i,j)=(1,2)$ or $(4,5)$), or in the terms with free indices, $P_{i,A_j}P_{3,A_i}-(i\leftrightarrow j)$. For the former, by shifting the Mellin variable $\gamma_{i(j)3}\to \gamma_{i(j)3}+1$ we have
\be
\label{mame}
P_i\cdot P_3-(i\leftrightarrow j)\to -\frac{1}{2}\gamma_{i3}-(i\leftrightarrow j)=\frac{1}{2}\sum_{k\neq i,3}\gamma_{ik}-(i\leftrightarrow j)\to -\sum_{k\neq i}P_i\cdot P_k-(i\leftrightarrow j)\ ,
\ee
where in the second step we have used $\Delta_i=\Delta_j=d-1$ and in the last step we have shifted $\gamma_{i(j)k}\to\gamma_{i(j)k}-1$. For the latter, we can use (\ref{eigD2}) to replace $P_{3,A_{i(j)}}$. So, it becomes
\be
\label{mame2}
P_{i,A_j}P_{3,A_i}-(i\leftrightarrow j)\to-P_{i,A_j}\sum_{k\neq 3}P_{k,A_i}-(i\leftrightarrow j)\ .
\ee

Combining (\ref{mame}) and (\ref{mame2}), the net result is that we can replace $P_3$ in (\ref{M5channel2}) by $-\sum_{k\neq 3}P_k$. After getting rid of the $P_3$ dependence, and interchanging the labels $1\leftrightarrow 2$ and $3\leftrightarrow 5$, we can read off that
\begin{multline}
\label{}
\tilde{\mathscr{M}}_{\texttt{3v3v3v},A_1A_2\cdots A_5}^{a_1a_2\cdots a_5}=-ig^3f^{a_1a_2b'}f^{a_5b''b'}f^{a_3a_4b''}\sum_{m',n=0}^\infty\Big\{X_{b''b',A_5}\Big\}\V_3^{m',n,0}\\
\times\frac{\Big\{X_{34}^{A_{b''}}\Big\}\V_3^{m',0,0}}{4m'!\Gamma\left(\frac{d}{2}+m'\right)\left(\gamma_{34}'-\frac{d}{2}+m'\right)}  \times \frac{\Big\{X_{12}^{A_{b'}}\Big\}\V_3^{n,0,0}}{4n!\Gamma\left(\frac{d}{2}+n\right)\left(\gamma_{12}'-\frac{d}{2}+n\right)}\ ,
\end{multline}
with $P_{b''}\equiv P_3+P_4\ , \ P_{b'}\equiv P_1+P_2$. Plugging it in (\ref{6Jchannel2}), we see that the poles are at
\begin{subequations}
\begin{equation}
\gamma_{56} = \frac{d}{2} - m, \quad \gamma_{12} = \frac{d}{2} - n - n_{12}, \quad \gamma_{34} = \frac{d}{2} - m' - n_{34}.
\end{equation}
And from the delta function restriction,
\be
\gamma_{13}+\gamma_{14}+\gamma_{23}+\gamma_{24}=&\frac{\sum_{i=1}^4\Delta_i-\Delta_5-\Delta_6+2}{2}+\gamma_{56}-\gamma_{12}-\gamma_{34}\\
=&\frac{d}{2}-m+n+m'+n_{12}+n_{34}\ .
\ee
The sum over $n_{ij}$ in (\ref{6Jchannel2}) at the poles yields
\be
\sum_{\sum_{i<j}^4n_{ij}=m}\prod_{i<j}^4\frac{(\gamma_{ij})_{n_{ij}}}{n_{ij}!}=&\sum_{n_{12}=0}^m\sum_{n_{34}=0}^{m-n_{12}}\frac{(\frac{d}{2}-n-n_{12})_{n_{12}}}{n_{12}!}\frac{(\frac{d}{2}-m'-n_{34})_{n_{34}}}{n_{34}!}\\
&\times\frac{(\frac{d}{2}-m+n+m'+n_{12}+n_{34})_{m-n_{12}-n_{34}}}{(m-n_{12}-n_{34})!}\ .
\ee
Shifts of $m'\to m'-n_{34}$ and $n\to n-n_{12}$ lead to
\begin{multline}
\sum_{n_{12}=0}^m\sum_{n_{34}=0}^{m-n_{12}}\frac{(\frac{d}{2}-n-n_{12})_{n_{12}}}{n_{12}!}\frac{(\frac{d}{2}-m'-n_{34})_{n_{34}}}{n_{34}!}\frac{(\frac{d}{2}-m+n+m'+n_{12}+n_{34})_{m-n_{12}-n_{34}}}{(m-n_{12}-n_{34})!}\\
\times\frac{\Gamma(d-1)(\frac{d}{2}-m')_{m'}}{4m'!\Gamma\left(\frac{d}{2}+m'\right)\left(\gamma_{34}+n_{34}-\frac{d}{2}+m'\right)}\V_3^{m',n,0}\frac{\Gamma(d-1)(\frac{d}{2}-n)_n}{4n!\Gamma\left(\frac{d}{2}+n\right)\left(\gamma_{12}+n_{12}-\frac{d}{2}+n\right)}\\
\to \frac{1}{m!}\frac{\Gamma(d-1)(\frac{d}{2}-m')_{m'}}{4m'!\Gamma\left(\frac{d}{2}+m'\right)\left(\gamma_{34}-\frac{d}{2}+m'\right)}\V_3^{m',n,m} \frac{\Gamma(d-1)(\frac{d}{2}-n)_n}{4n!\Gamma\left(\frac{d}{2}+n\right)\left(\gamma_{12}-\frac{d}{2}+n\right)}\ ,
\end{multline}
where we have defined
\begin{multline}
\label{V33}
\V_3^{m',n,m}\equiv \sum_{n_{12}=0}^{\min\{m,n\}}\sum_{n_{34}=0}^{\min\{m-n_{12},m'\}}\frac{m!}{n_{12}!n_{34}!}\frac{\left(\frac{d}{2}-m+n+m'\right)_{m-n_{12}-n_{34}}}{(m-n_{12}-n_{34})!}(m'-n_{34}+1)_{n_{34}}\\\times\left(\frac{d}{2}+m'-n_{34}\right)_{n_{34}}(n-n_{12}+1)_{n_{12}}\left(\frac{d}{2}+n-n_{12}\right)_{n_{12}}\V_3^{m'-n_{34},n-n_{12},0}\ .
\end{multline}
\end{subequations}
Presumably this definition is symmetric among $m',n$ and $m$, and reduces to the one defined in (\ref{V3mn0}) when one of the integer is set to 0. While we don't have a proof for the symmetry property, we can check that the reduction property is true. First, plugging $m=0$ in (\ref{V33}), we find that the r.h.s reduces to $\V_3^{m',n,0}$, consistent with the l.h.s. Besides, from the definition (\ref{V33}), we find $\V_3^{m',0,m}=\V_3^{0,m',m}$, and it is equal to $\V_3^{m,m',0}$ from the definition in (\ref{V3mn0}).

With the aid of the newly defined $\V_3^{m',n,m}$, we can rewrite the Mellin amplitude (\ref{6Jchannel2}) as
\begin{eBox}
\begin{multline}
\label{M6channel2}
\mathscr{M}_{\texttt{Snowflake}}^{M_1M_2\cdots M_6}
=g^4f^{a_1a_2b'}f^{a_3a_4b''}f^{a_5a_6b}f^{bb''b'}\frac{\pi^h}{2} \left(\Di{6} \right) \\\sum_{m',n,m=0}^\infty  \Big\{X_{b''b',M}\Big\}\V_3^{m',n,m}\times\frac{\Big\{X_{56}^M\Big\}\V_3^{m,0,0}}{4m!\Gamma\left(\frac{d}{2}+m\right)\left(\gamma_{56}-\frac{d}{2}+m\right)}\\\times\frac{\Big\{X_{34}^{A_{b''}}\Big\}\V_3^{m',0,0}}{4m'!\Gamma\left(\frac{d}{2}+m'\right)\left(\gamma_{34}-\frac{d}{2}+m'\right)}\times \frac{\Big\{X_{12}^{A_{b'}}\Big\}\V_3^{n,0,0}}{4n!\Gamma\left(\frac{d}{2}+n\right)\left(\gamma_{12}-\frac{d}{2}+n\right)}\ .
\end{multline}
\end{eBox}
Compare it with the flat-space analog
\small
\begin{multline}
\label{}
\mathscr{A}_{\texttt{Snowflake}}^{M_1M_2\cdots M_6}=g^4f^{a_1a_2b'}f^{a_3a_4b''}f^{a_5a_6b}f^{bb''b'}\left(\eta_{M''M'}q''_M-2\eta_{M''M}q''_{M'}-(q''M''\leftrightarrow q'M')\right)\\
\times \left(\eta_{A_5A_6}k_5^M-2\eta_{A_5}^Mk_{5,A_6}-(5\leftrightarrow 6)\right)\frac{i}{(k_5+k_6)^2}\left(\eta_{A_3A_4}k_3^{M''}-2\eta_{A_3}^{M''}k_{3,A_4}-(3\leftrightarrow 4)\right) \\
\times\frac{i}{(k_3+k_4)^2}\left(\eta_{A_1A_2}k_1^{M'}-2\eta_{A_1}^{M'}k_{1,A_2}-(1\leftrightarrow 2)\right)\frac{i}{(k_1+k_2)^2}\ ,
\end{multline}
\normalsize
where $q''\equiv k_3+k_4$ and $q'\equiv k_1+k_2$, we have the dictionary that $ik_i\to 2P_i$ for the momenta,
\begin{equation}
\frac{i}{(k_i+k_j)^2}\to \frac{1}{4m!\Gamma\left(\frac{d}{2}+m\right)\left(\gamma_{ij}-\frac{d}{2}+m\right)}\ ,
\end{equation}
with $m$ to be summed over for the propagators, and $g\to g\ \V_3^{m',n,m}$ for the three-vertex coupling constant.

\subsection{Six-point amplitude: channel with two three-vertices and a four-vertex}

\begin{figure}
	\centering
\includegraphics[scale=1.2]{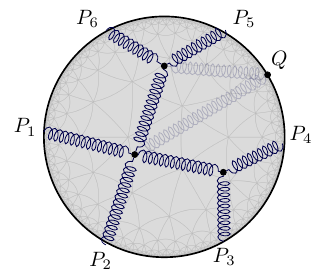}
\caption{\label{6channel3} The channel of six point gluon amplitude for (\ref{6Jchannel3})}
\end{figure}

Now we calculate the six-point gluon Mellin amplitude in another channel as depicted in Figure \ref{6channel3}. The amplitude can be factorized into a five-point amplitude and a three-point amplitude.
Using (\ref{eq:3J}), (\ref{M5channel1}) and (\ref{nM}), we have
\footnotesize
\begin{multline}
\label{6Jchannel3}
\mathscr{M}_{\texttt{3v3v4v}}^{M_1M_2\cdots M_6}
=ig^4\frac{\pi^h}{2} \left(\Di{6} \right)\sum_{m=0}^\infty\frac{\V_3^{m,0,0}}{4\Gamma\left(\frac{d}{2}+m\right)(\gamma_{56}-\frac{d}{2}+m)}\\
\times\sum_{\sum_{i<j}^4n_{ij}=m}\prod_{i<j}^4\frac{(\gamma_{ij})_{n_{ij}}}{n_{ij}!} \sum_{n=0}^\infty\frac{\V_3^{n,0,0}\times \V_4^{n,0,0,0}}{4n!\Gamma\left(\frac{d}{2}+n\right)(\gamma_{34}+n_{34}-\frac{d}{2}+n)}f^{a_5a_6b}\Big\{X_{56}^M\Big\}«f^{a_3a_4b''}\\
\times\left(\left(f^{a_1b''b'}f^{a_2bb'}+f^{a_2b''b'}f^{a_1bb'}\right)\Big\{X_{34,M}\Big\}\eta_{A_1A_2}+\text{cyclic perm. of }(A_1a_1,A_2a_2,Mb)\right)\ .
\end{multline}
\normalsize
\begin{subequations}
The poles of Mellin variables are at
\be
\gamma_{56}=\frac{d}{2}-m,\quad \gamma_{34}=\frac{d}{2}-n-n_{34}\ .
\ee
Then, from the equation which results from the restrictions on the Mellin variables, i.e.,
\be
\Delta_5+\Delta_6+1-2\gamma_{56}=\sum_{i=1}^4\Delta_i+1-2\sum_{i<j}^4\gamma_{ij}\ ,
\ee
we have
\be
\sum_{\sum_{i<j}^4n_{ij}=m}\prod_{i<j}^4\frac{(\gamma_{ij})_{n_{ij}}}{n_{ij}!}=\sum_{n_{34}=0}^m\frac{(\frac{d}{2}-n-n_{34})_{n_{34}}(d-1-m+n+n_{34})_{m-n_{34}}}{n_{34}!(m-n_{34})!}\ .
\ee
A shift of $n\to n-n_{34}$ amounts to
\be
&\sum_{n_{34}=0}^m\frac{(\frac{d}{2}-n-n_{34})_{n_{34}}(d-1-m+n+n_{34})_{m-n_{34}}}{n_{34}!(m-n_{34})!}\frac{\left(\frac{d}{2}-n\right)_n(d-1-n)_n}{n!\Gamma\left(\frac{d}{2}+n\right)(\gamma_{34}+n_{34}-\frac{d}{2}+n)}\\
\to&\frac{\V_4^{m,n,0,0}}{\Gamma\left(\frac{3d-4}{2}\right)}\frac{\left(\frac{d}{2}-n\right)_n}{m!n!\Gamma\left(\frac{d}{2}+n\right)(\gamma_{34}-\frac{d}{2}+n)}\ ,
\ee
where
\be
\label{V4mn00}
\V_4^{m,n,0,0}\equiv &\Gamma\left(\frac{3d-4}{2}\right)m!\sum_{n_{34}=0}^{\min\{m,n\}}\frac{(d-1-m+n)_{m-n_{34}}}{n_{34}!(m-n_{34})!}\\
&\times(n-n_{34}+1)_{n_{34}}\left(\frac{d}{2}+n-n_{34}\right)_{n_{34}}(d-1-n+n_{34})_{n-n_{34}}\ .
\ee
\end{subequations}
One can check that from this definition $\V_4^{m,0,0,0}=\V_4^{0,m,0,0}$, which is also identical to the $\V_4^{m,0,0,0}$ given in (\ref{V4m000}). Now, we can rewrite (\ref{6Jchannel3}) as
\begin{eBox}
\begin{multline}
\label{M6channel3}
\mathscr{M}_{\texttt{3v3v4v}}^{M_1M_2\cdots M_6}
=ig^4\frac{\pi^h}{2} \left(\Di{6} \right)   \sum_{m,n=0}^\infty \V_4^{m,n,0,0}
\times\frac{\Big\{X_{56}^M\Big\}\V_3^{m,0,0}}{4m!\Gamma\left(\frac{d}{2}+m\right)(\gamma_{56}-\frac{d}{2}+m)}\\
\times \frac{\Big\{X_{34}^{M''}\Big\}\V_3^{n,0,0}}{4n!\Gamma\left(\frac{d}{2}+n\right)(\gamma_{34}-\frac{d}{2}+n)}f^{a_5a_6b}f^{a_3a_4b''} \left(\left(f^{a_1b''b'}f^{a_2bb'}+f^{a_2b''b'}f^{a_1bb'}\right)\right.\\ 
\times\left.\eta_{A_1A_2}\eta_{MM''} + \text{cyclic perm. of }(A_1a_1,A_2a_2,Mb)\right)\ .
\end{multline}
\end{eBox}
Like the previous cases, the Mellin amplitude (\ref{M6channel3}) can be obtained from the flat-space Feynman rules by replacing $ik_i\to 2P_i$ for the momenta,
\begin{equation}
\frac{i}{(k_i+k_j)^2}\to \frac{1}{4m!\Gamma\left(\frac{d}{2}+m\right)\left(\gamma_{ij}-\frac{d}{2}+m\right)}\ ,
\end{equation}
with $m$ to be summed over for the propagators, $g\to g\ \V_3^{m,0,0}$ for the three-vertex coupling constant and $g^2\to g^2\ \V_4^{m,n,0,0}$ for the four-vertex coupling constant.
\subsection{Seven-point amplitude: scarecrow channel}
\begin{figure}
	\centering
\includegraphics[scale=1.2]{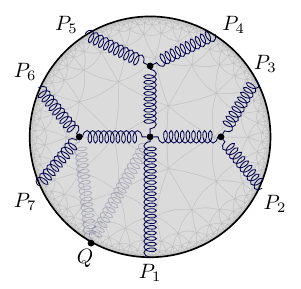}
\caption{\label{7channel1} The "scarecrow" channel of seven-point gluon amplitude for (\ref{7Jchannel1})}
\end{figure}
Let's proceed to compute the seven-point gluon amplitude, as depicted in Figure \ref{7channel1}. Using (\ref{eq:3J}), (\ref{M6channel3}) and (\ref{nM}), we have
\footnotesize
\begin{multline}
\label{7Jchannel1}
\mathscr{M}_{\texttt{Scarecrow}}^{M_1M_2\cdots M_7}
=-g^5\frac{\pi^h}{2} \left(\Di{7} \right)\sum_{m=0}^\infty\frac{\Gamma\left(d-1\right)\left(\frac{d}{2}-m\right)_m}{4\Gamma\left(\frac{d}{2}+m\right)(\gamma_{67}-\frac{d}{2}+m)} \\
\times\sum_{\sum_{i<j}^5n_{ij}=m}\prod_{i<j}^5\frac{(\gamma_{ij})_{n_{ij}}}{n_{ij}!} \sum_{m',n=0}^\infty \V_4^{m',n,0,0}\frac{\Gamma\left(d-1\right)\left(\frac{d}{2}-m'\right)_{m'}}{4m'!\Gamma\left(\frac{d}{2}+m'\right)(\gamma_{45}+n_{45}-\frac{d}{2}+m')}\\
\times\frac{\Gamma\left(d-1\right)\left(\frac{d}{2}-n\right)_n}{4n!\Gamma\left(\frac{d}{2}+n\right)(\gamma_{23}+n_{23}-\frac{d}{2}+n)}f^{ba_6a_7}\Big\{X_{67}^M\Big\}f^{a_4a_5c}\Big\{X_{45}^N\Big\}f^{a_2a_3b''}\\
\times\left(\left(f^{bb''b'}f^{a_1cb'}+f^{a_1b''b'}f^{bcb'}\right)\Big\{X_{23,N}\Big\}\eta_{MA_1} +\text{cyclic perm. of }(Mb,A_1a_1,Nc)\right)\ .
\end{multline}
\normalsize
At the pole,
\begin{subequations}
\be
\gamma_{67}=\frac{d}{2}-m,\quad \gamma_{23}=\frac{d}{2}-n-n_{23},\quad \gamma_{45}=\frac{d}{2}-m'-n_{45}\ ,
\ee
with
$\Delta_6+\Delta_7+1-2\gamma_{67}=\sum_{i=1}^5\Delta_i+2-2\sum_{i<j}^5\gamma_{ij}$, we have
\be
&\sum_{\sum_{i<j}^5n_{ij}=m}\prod_{i<j}^5\frac{(\gamma_{ij})_{n_{ij}}}{n_{ij}!}\\
=&\sum_{n_{23}=0}^m\sum_{n_{45}=0}^{m-n_{23}}\frac{(\frac{d}{2}-n-n_{23})_{n_{23}}(\frac{d}{2}-m'-n_{45})_{n_{45}}(d-1-m+n+m'+n_{23}+n_{45})_{m-n_{23}-n_{45}}}{n_{23}!n_{45}!(m-n_{23}-n_{45})!}\ .
\ee
Shift $n\to n-n_{23}$ and $m'\to m'-n_{45}$
\begin{multline}
\sum_{n_{23}=0}^m\sum_{n_{45}=0}^{m-n_{23}}\frac{(\frac{d}{2}-n-n_{23})_{n_{23}}(\frac{d}{2}-m'-n_{45})_{n_{45}}(d-1-m+n+m'+n_{23}+n_{45})_{m-n_{23}-n_{45}}}{n_{23}!n_{45}!(m-n_{23}-n_{45})!}\\
\times\frac{\left(\frac{d}{2}-m'\right)_{m'}}{m'!\Gamma\left(\frac{d}{2}+m'\right)(\gamma_{45}+n_{45}-\frac{d}{2}+m')}\V_4^{m',n,0,0}\frac{\left(\frac{d}{2}-n\right)_n}{n!\Gamma\left(\frac{d}{2}+n\right)(\gamma_{23}+n_{23}-\frac{d}{2}+n)}\\
\to \V_4^{m',n,m,0}\frac{1}{m!}\frac{\left(\frac{d}{2}-m'\right)_{m'}}{m'!\Gamma\left(\frac{d}{2}+m'\right)(\gamma_{45}-\frac{d}{2}+m')}\frac{\left(\frac{d}{2}-n\right)_n}{n!\Gamma\left(\frac{d}{2}+n\right)(\gamma_{23}-\frac{d}{2}+n)}\ ,
\end{multline}
where
\begin{multline}
\label{V4mnm0}
\V_4^{m',n,m,0}\equiv m!\sum_{n_{23}=0}^{\min\{m,n\}}\sum_{n_{45}=0}^{\min\{m-n_{23},m'\}}\frac{(d-1-m+n+m')_{m-n_{23}-n_{45}}}{n_{23}!n_{45}!(m-n_{23}-n_{45})!}(m'-n_{45}+1)_{n_{45}} \\
\times \left(\frac{d}{2}+m'-n_{45}\right)_{n_{45}}\V_4^{m'-n_{45},n-n_{23},0,0}
\times(n-n_{23}+1)_{n_{23}}\left(\frac{d}{2}+n-n_{23}\right)_{n_{23}}\ .
\end{multline}
\end{subequations}

One can check that from this definition $\V_4^{m,n,0,0}=\V_4^{0,m,n,0}=\V_4^{m,0,n,0}$ and reduces to $\V_4^{m,n,0,0}$ in (\ref{V4mn00}). Furthermore, with (\ref{V4mnm0}), we can rewrite the Mellin amplitude (\ref{7Jchannel1}) as
\begin{eBox}
\begin{multline}
\label{M7channel1}
\mathscr{M}_{\texttt{Scarecrow}}^{M_1M_2\cdots M_7}
=-g^5\frac{\pi^h}{2} \left(\Di{7} \right)   \sum_{m,n,m'=0}^\infty \V_4^{m',n,m,0} \frac{\Big\{X_{67}^M\Big\}\V_3^{m,0,0}}{4m!\Gamma\left(\frac{d}{2}+m\right)(\gamma_{67}-\frac{d}{2}+m)}\\
\times\frac{\Big\{X_{45}^{M'}\Big\}\V_3^{m',0,0}}{4m'!\Gamma\left(\frac{d}{2}+m'\right)(\gamma_{45}-\frac{d}{2}+m')} \times \frac{\Big\{X_{23}^N\Big\}\V_3^{n,0,0}}{4n!\Gamma\left(\frac{d}{2}+n\right)(\gamma_{23}-\frac{d}{2}+n) }f^{a_6a_7b}f^{a_4a_5c}f^{a_2a_3b''}\\
\times\left(\left(f^{bb''b'}f^{a_1cb'}+f^{a_1b''b'}f^{bcb'}\right)\eta_{MA_1}\eta_{M'N}+\text{cyclic perm. of }(Mb,A_1a_1,M'c)\right)\ ,
\end{multline}
\end{eBox}
which can be mapped from the flat-space counterpart by replacing $ik_i\to 2P_i$ for the momenta,
\begin{equation}
\frac{i}{(k_i+k_j)^2}\to \frac{1}{4m!\Gamma\left(\frac{d}{2}+m\right)\left(\gamma_{ij}-\frac{d}{2}+m\right)}\ ,
\end{equation}
with $m$ to be summed over for the propagators, $g\to g\ \V_3^{m,0,0}$ for the three-vertex coupling constant and $g^2\to g^2\ \V_4^{m',n,m,0}$ for the four-vertex coupling constant.
\subsection{Eight-point amplitude: drone channel}
\begin{figure}
	\centering
\includegraphics[scale=1.2]{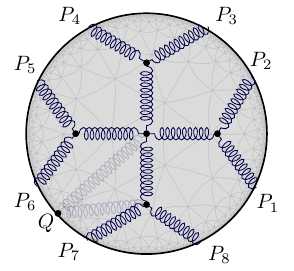}
\caption{\label{8channel1} The ``drone'' channel of eight-point gluon amplitude for (\ref{8Jchannel1})}
\end{figure}
In this subsection we calculate the eight-point gluon amplitude in the drone channel. As shown in Figure \ref{8channel1}. To calculate the amplitude, we factorize the diagram into a seven-point amplitude and a three-point amplitude. Using (\ref{eq:3J}), (\ref{M7channel1}) and (\ref{nM}), we have
\footnotesize
\begin{multline}
\label{8Jchannel1}
\mathscr{M}_{\texttt{Drone}}^{M_1M_2\cdots M_8}
=-ig^6\frac{\pi^h}{2} \left(\Di{8} \right) \sum_{m=0}^\infty\frac{\Gamma\left(d-1\right)\left(\frac{d}{2}-m\right)_m}{4m!\Gamma\left(\frac{d}{2}+m\right)(\gamma_{78}-\frac{d}{2}+m)}\sum_{\sum_{i<j}^6n_{ij}=m}\prod_{i<j}^6\frac{(\gamma_{ij})_{n_{ij}}}{n_{ij}!}\\
\sum_{n',n,m'=0}^\infty \V_4^{m',n,n',0}\frac{\Gamma\left(d-1\right)\left(\frac{d}{2}-n'\right)_{n'}}{4n'!\Gamma\left(\frac{d}{2}+n'\right)(\gamma_{56}+n_{56}-\frac{d}{2}+n')}\frac{\Gamma\left(d-1\right)\left(\frac{d}{2}-m'\right)_{m'}}{4m'!\Gamma\left(\frac{d}{2}+m'\right)(\gamma_{34}+n_{34}-\frac{d}{2}+m')} \\
\times\frac{\Gamma\left(d-1\right)\left(\frac{d}{2}-n\right)_n}{4n!\Gamma\left(\frac{d}{2}+n\right)(\gamma_{12}+n_{12}-\frac{d}{2}+n)}f^{a_7a_8b}\Big\{X_{78}^M\Big\} f^{a_5a_6c'}\Big\{X_{56}^{M'}\Big\} f^{a_3a_4c}\Big\{X_{34}^{N}\Big\}\\
\times \left(\left(f^{c'b''b'}f^{bcb'}+f^{bb''b'}f^{c'cb'}\right)\Big\{X_{12,N}\Big\}\eta_{MM'} +\text{cyclic perm. of }(M'c',Mb,Nc)\right)\ .
\end{multline}
\normalsize
\begin{subequations}
At the pole,
\be
\gamma_{78}=\frac{d}{2}-m,\quad \gamma_{12}=\frac{d}{2}-n-n_{12},\quad \gamma_{34}=\frac{d}{2}-m'-n_{34},\quad \gamma_{56}=\frac{d}{2}-n''-n_{56}\ ,
\ee
with
\be
\Delta_7+\Delta_8+1-2\gamma_{78}=\sum_{i=1}^6\Delta_i+3-2\sum_{i<j}^6\gamma_{ij}\ ,
\ee
we have
\be
&\sum_{\sum_{i<j}^6n_{ij}=m}\prod_{i<j}^6\frac{(\gamma_{ij})_{n_{ij}}}{n_{ij}!}\\
=&\sum_{n_{12}=0}^m\sum_{n_{34}=0}^{m-n_{12}}\sum_{n_{56}=0}^{m-n_{12}-n_{34}}\frac{(\frac{d}{2}-n-n_{12})_{n_{12}}(\frac{d}{2}-m'-n_{34})_{n_{34}}(\frac{d}{2}-n'-n_{56})_{n_{56}}}{n_{12}!n_{34}!n_{56}!}\\
&\times\frac{(d-1-m+n+m'+n'+n_{12}+n_{34}+n_{56})_{m-n_{12}-n_{34}-n_{56}}}{(m-n_{12}-n_{34}-n_{56})!}\ .
\ee
Shift $n\to n-n_{12}$, $m'\to m'-n_{34}$ and $n'\to n'-n_{56}$
\footnotesize
\begin{multline}
\left(\sum_{n_{12}=0}^m\sum_{n_{34}=0}^{m-n_{12}}\sum_{n_{56}=0}^{m-n_{12}-n_{34}}\frac{(\frac{d}{2}-n-n_{12})_{n_{12}}(\frac{d}{2}-m'-n_{34})_{n_{34}}(\frac{d}{2}-n'-n_{56})_{n_{56}}}{n_{12}!n_{34}!n_{56}!}\right.\\
\times\frac{(d-1-m+n+m'+n'+n_{12}+n_{34}+n_{56})_{m-n_{12}-n_{34}-n_{56}}}{(m-n_{12}-n_{34}-n_{56})!}\V_4^{m',n,n',0}\frac{\left(\frac{d}{2}-n'\right)_{n'}}{n'!\Gamma\left(\frac{d}{2}+n'\right)(\gamma_{56}+n_{56}-\frac{d}{2}+n')}\\\times\left.\frac{\left(\frac{d}{2}-m'\right)_{m'}}{m'!\Gamma\left(\frac{d}{2}+m'\right)(\gamma_{34}+n_{34}-\frac{d}{2}+m')}\frac{\left(\frac{d}{2}-n\right)_n}{n!\Gamma\left(\frac{d}{2}+n\right)(\gamma_{12}+n_{12}-\frac{d}{2}+n)}\right)\\
\to \left(\V_4^{m',n,n',m}\frac{\left(\frac{d}{2}-n'\right)_{n'}}{n'!\Gamma\left(\frac{d}{2}+n'\right)(\gamma_{56}-\frac{d}{2}+n')}\frac{\left(\frac{d}{2}-m'\right)_{m'}}{m'!\Gamma\left(\frac{d}{2}+m'\right)(\gamma_{34}-\frac{d}{2}+m')}\frac{\left(\frac{d}{2}-n\right)_n}{n!\Gamma\left(\frac{d}{2}+n\right)(\gamma_{12}-\frac{d}{2}+n)}\right)\ ,
\end{multline}
\normalsize
where
\begin{multline}
\label{V4mnnm}
\V_4^{m',n,n',m}\\
\equiv m!\sum_{n_{12}=0}^{\min\{m,n\}}\sum_{n_{34}=0}^{\min\{m-n_{34},m'\}}\sum_{n_{56}=0}^{\min\{m-n_{12}-n_{34},n'\}}\frac{(d-1-m+n+m'+n')_{m-n_{12}-n_{34}-n_{56}}}{n_{12}!n_{34}!n_{56}!(m-n_{12}-n_{34}-n_{56})!}\\
\times\V_4^{m'-n_{34},n-n_{12},n'-n_{56},0}(n'-n_{56}+1)_{n_{56}}\left(\frac{d}{2}+n'-n_{56}\right)_{n_{56}}\\
\times(m'-n_{34}+1)_{n_{34}}\left(\frac{d}{2}+m'-n_{34}\right)_{n_{34}}(n-n_{12}+1)_{n_{12}}\left(\frac{d}{2}+n-n_{12}\right)_{n_{12}}\ .
\end{multline}
\end{subequations}
One can check that from this definition $\V_4^{m',n,m,0}=\V_4^{m',n,0,m}=\V_4^{m',0,n,m}=\V_4^{0,m',n,m}$ and reduces to $\V_4^{m',n,m,0}$ in (\ref{V4mnm0}). With such a definition of $\V_4^{m',n,n',m}$, we can rewrite the Mellin amplitude (\ref{8Jchannel1}) as
\begin{eBox}
\begin{multline}
\label{M8channel1}
\mathscr{M}_{\texttt{Drone}}^{M_1M_2\cdots M_8}
=-ig^6\frac{\pi^h}{2} \left(\Di{8} \right) \sum_{n',n,m',m=0}^\infty \V_4^{m',n,n',m}\frac{\Big\{X_{78}^M \Big\}\V_3^{m,0,0}}{4m!\Gamma\left(\frac{d}{2}+m\right)(\gamma_{78}-\frac{d}{2}+m)}\\
\times\frac{\Big\{X_{56}^{M'}\Big\}\V_3^{n',0,0}}{4n'!\Gamma\left(\frac{d}{2}+n'\right)(\gamma_{56}-\frac{d}{2}+n')}\times \frac{\Big\{X_{34}^{N}\Big\}\V_3^{m',0,0}}{4m'!\Gamma\left(\frac{d}{2}+m'\right)(\gamma_{34}-\frac{d}{2}+m')}\\
\times\frac{\Big\{X_{12}^{M''}\Big\}\V_3^{n,0,0}}{4n!\Gamma\left(\frac{d}{2}+n\right)(\gamma_{12}-\frac{d}{2}+n)} f^{a_7a_8b}f^{a_5a_6c'} f^{a_3a_4c}f^{a_1a_2b''}\\
\times \left(\left(f^{c'b''b'}f^{bcb'}+f^{bb''b'}f^{c'cb'}\right)\eta_{MM'}\eta_{NM''} +\text{cyclic perm. of }(M'c',Mb,Nc)\right)\ ,
\end{multline}
\end{eBox}
which is related to the flat-space counterpart by the following replacements up to an overall $\frac{\pi^h}{2}\Di{8}$. For the momenta, $ik_i\to 2P_i$. For propagators,
\begin{equation}
\frac{i}{(k_i+k_j)^2}\to \frac{1}{4m!\Gamma\left(\frac{d}{2}+m\right)\left(\gamma_{ij}-\frac{d}{2}+m\right)}\ ,
\end{equation}
with $m$ to be summed over. For the three-vertex coupling constant, $g\to g\ \V_3^{m,0,0}$, and for the four-vertex coupling constant $g^2\to g^2\ \V_4^{m',n,n',m}$.
\subsection{Dictionary between gluon Mellin amplitude and flat-space gluon amplitude}
Reflecting on the diverse range of examples that we have provided, spanning from three to eight-point amplitudes, an intriguing similarity emerges between the Mellin amplitude in Anti-de Sitter (AdS) spaces and the flat-space amplitude perturbatively derived from the Feynman rules. This correspondence is not just supeficial and there is a precise dictionary between them, as shown in Table \ref{dict}.

The emergence of kinematic variables in scattering amplitudes is traced back to the derivative terms in the action. Specifically, in flat space, applying the derivative $ \partial_{x^A} $ to the Fourier basis $e^{ik_ix}$  introduces a factor, $ik^A$. In contrast, within AdS space, the operation of $\partial_{X^A}$ on the boundary-to-bulk propagator, represented as $\mathcal{E}_{\Delta_i}^{M_iA_i}(P_i,X)$, produces a factor of $2P_i^A$. This analogy provides a rationale for the presence of $ik$ and $2P_i$ on respective sides of the established dictionary.

Indeed, the map $2P_i \leftrightarrow ik_i$ has been substantiated through several examples, which demonstrate a notable parallelism in their behaviors. Firstly, both Mellin amplitudes and flat space amplitudes conform to the null condition, expressed as $P^2 = 0$ and $k^2 = 0$. Additionally, analogues of momentum conservation, referenced by equations \eqref{eigD1} and \eqref{eigD2}, are observed for these boundary points.

For the internal propagator in the flat-space amplitude, i.e. $i/(\sum_ik_i)^2=i/2\sum_{i<j}k_i\cdot k_j$, we have $1/\tilde{\sum}_{i<j}\gamma_{ij}$ on the Mellin side, where we have defined
\begin{equation}
\label{propagator}
\tilde{\sum_{i<j}}\gamma_{ij}\equiv 4m! ~~ \Gamma\left(\frac{d}{2} + m\right) \left[\sum_{i<j} \gamma_{ij} + \frac{(d-1)}{2} - \frac{1}{2} \sum_i (d - 1 - \delta_i) + m\right],
\end{equation}
where again $\delta_i$ denotes the scaling dimension of $P_i$ in the Mellin amplitude. The map between the Mandelstam invariants, $k_i \cdot k_j$ and $\gamma_{ij}$ resembles the relationship in the scalar scenario, as referenced in \cite{Paulos:2011ie}. 


From (\ref{propagator}), it is clear that each bulk-to-bulk propagator is associated with an integer $m$ to be summed over from 0 to $\infty$. Then, for a three-vertex connecting the propagators associated with integers $m_1,m_2$ and $m_3$, the three-vertex coupling is
\be
g\leftrightarrow g\ \V_3^{m_1,m_2,m_3}\ .
\ee
This may contain boundary-to-bulk propagators with $m_i=0$. Similarly, for a four-vertex connecting propagators associated with integers $m_1,m_2,m_3$ and $m_4$, the coupling is supposed to be
\be
g^2\leftrightarrow g^2\ \V_4^{m_1,m_2,m_3,m_4}\ .
\ee

It is also noteworthy that in scalar cases a high energy limit $\gamma_{ij}\to\infty$ takes Mellin amplitudes to flat-space amplitudes up to a transform~\cite{Fitzpatrick:2011ia}. Specifically, as reviewed in Appendix \ref{fllim}, in the flat-space limit an $n$-point scalar Mellin amplitude becomes
\begin{equation}
\begin{split}
\mathscr{M}_n(\Delta_i,\gamma_{ij})\approx \frac{\pi^h}{2}\prod_{i=1}^n\frac{C_{\Delta_i}}{\Gamma(\Delta_i)}\int_0^\infty d\beta\beta^{\frac{1}{2}(\sum_{i=1}^n\Delta_i-d)-1}e^{-\beta}\mathscr{A}_n(p_i\cdot p_j=2\beta\gamma_{ij})\ .
\end{split}
\end{equation}
It is natural to generalize the flat-space limit to gluon Mellin amplitudes. Following the dictionary Table \ref{dict}, we propose that the generalization of flat-space limit is
\begin{equation}
\begin{split}
\mathscr{M}_n^{M_1\cdots M_n}\approx &\frac{\pi^h}{2}\left(\Di{n}\right)\int_0^\infty d\beta\beta^{\frac{\sum_{i=1}^n\Delta_i-d}{2}-1}e^{-\beta} \\
&\times\mathscr{A}_{n,A_1A_2\cdots A_n} \bigg(\frac{i}{\big(\sum_i k_i\big)^2}\to \frac{1}{4\beta\sum_{i<j}\gamma_{ij}},ik_i\to 2\sqrt{\beta} P_i\bigg)\ .
\end{split}
\end{equation}
We perform several checks for this formula in Appendix \ref{fllimgl}.

\section{Conclusion and outlook}\label{section5}
In this study, we present a rigorous computation of the gluon amplitude in Anti-de Sitter (AdS) space. We employed the embedding formalism, Mellin space techniques, and an approach utilizing differential operators, successfully computing novel higher-point correlators. Despite the complexity of the intermediate steps, we distilled the Mellin results into remarkably succinct expressions. Intriguingly, we observed that our results exhibit a striking resemblance to structures in flat space. This work potentially opens new avenues for research.

It is conceivable that one could rewrite these expressions in spinor helicity formalism. Conventionally, flat space scattering amplitudes are written using polarization vectors. However, the preceding twenty years have witnessed advancements through the adoption of spinor helicity variables. These variables are distinguished by their transformation properties under the spinor representations of both the Lorentz group and the little group. It would be interesting to see if one could use a different variable to simplify the expressions.\footnote{For a bispinor formalism of AdS correlators in embedding space, see \cite{Binder:2020raz}}  

In the context of Mellin space, the domain of external gravitons presents a significant avenue for further exploration. Only a handful of works have been conducted in this direction \cite{Bonifacio:2022vwa, Raju:2012zs, Albayrak:2019yve,Albayrak:2023jzl}. The four-point external graviton falls within our investigative purview. Moreover, the potential to establish a map between flat space graviton scattering and graviton bulk scattering in the AdS framework is both promising and of considerable practical relevance from AdS as well as dS point of view.

Related to the graviton Mellin amplitudes, several research groups have made progress in the (A)dS color kinematics and double copy frontier, though predominantly limited to three and four-point configurations \cite{Armstrong:2020woi,Albayrak:2020fyp,Alday:2021odx,Diwakar:2021juk,Sivaramakrishnan:2021srm,Cheung:2022pdk,Herderschee:2022ntr,Farrow:2018yni,Lipstein:2019mpu,Jain:2021qcl,Zhou:2021gnu,Lee:2022fgr,Bissi:2022wuh,Armstrong:2023phb,Alday:2022lkk, Liang:2023zxo}. To truly harness the potential for double copy, it is important to systematically explore higher-point configurations! Digging deeper into these intricate structures not only broadens the range of computable amplitudes but also underscores the efficacy and profound insights offered by the double copy approach. Our method for tackling higher-point configurations and its resemblance to flat space could be important for constructing the double copy that mirrors the flat-space version.

An exciting opportunity presents itself in exploring the computation of spinning loops within AdS (see some work in this direction \cite{Giombi:2017hpr, Albayrak:2020bso, Huang:2023ppy}). Previous investigations into loop calculations in flat space have revealed crucial links between trees and loops, bridging gravitational theories with gauge theories. In flat space, these loop amplitudes also exhibit connections to intricate geometric structures. It would be fascinating to investigate whether similar patterns or connections emerge in the loops within AdS.

\section*{Acknowledgement}
The authors express sincere thanks for the enlightening discussions with Soner Albayrak. We thank Xinkang Wang for comment on the draft. We thank developers of Tikz, which was used in drawing the diagrams in this paper.

\appendix
\section{Three-point scalar and gluon amplitudes: Schwinger trick}
\label{3ptappendix}
In this appendix, we review the calculation of the three-point AdS amplitudes~\cite{Paulos:2011ie}. For the scalar case, as discussed above, the correlation function can be calculated by integrating the bulk-to-boundary propagators, i.e.
\be
\left\langle \mathcal{O}_1(P_1)\mathcal{O}_2(P_2)\mathcal{O}_3(P_3)\right\rangle=ig\int_{\text{AdS}}dX\ \mathcal{E}(P_1,X)~ \mathcal{E}(P_2,X) \mathcal{E}(P_3,X)\ .
\ee
To perform the integration, it is convenient to express the bulk-to-boundary propagator with Schwinger parameter, 
\be
\mathcal{E}(P_i,X)=\frac{C_{\Delta_i}}{\Gamma(\Delta_i)}\int_0^{+\infty}\frac{dt_i}{t_i}t_i^{\Delta_i} e^{2t_iP_i\cdot X}\ .
\ee
Then, the integration becomes
\be
\left\langle \mathcal{O}_1(P_1)\mathcal{O}_2(P_2)\mathcal{O}_3(P_3)\right\rangle=ig\prod_{i=1}^3\frac{C_{\Delta_i}}{\Gamma(\Delta_i)} \int_0^{+\infty}\frac{dt_i}{t_i}t^{\Delta_i}\int_{\text{AdS}}dX e^{2T\cdot X}\ ,
\ee
where $T\equiv \sum_{i=1}^3t_iP_i$. Since $t_i$ are positive and $P_i$ are null vectors, $T$ must be time-like. In the rest frame where $T=(T^0,T^\mu)=(|T|,0)$, parametrize the AdS$_{d+1}$ space by
\be
X=\left(\frac{1+x_0^2+x^2}{2x_0},\frac{1-x_0^2-x^2}{2x_0},x^\mu\right)\ ,
\ee
which satisfies the equation (\ref{AdS}) for $R=1$. Then,
\be
\int_{\text{AdS}}dX e^{2T\cdot X}=&\int_0^{+\infty}\frac{dx_0}{x_0^{d+1}}\int_0^{+\infty}d^dxe^{-\frac{1+x_0^2+x^2}{x_0}|T|}\\
=&\pi^h\int_0^{+\infty}\frac{dx_0}{x_0^{h+1}}e^{-x_0+\frac{T^2}{x_0}}\ .
\ee
So,
\be
\left\langle \mathcal{O}_1(P_1)\mathcal{O}_2(P_2)\mathcal{O}_3(P_3)\right\rangle=&ig\pi^h\prod_{i=1}^3\frac{C_{\Delta_i}}{\Gamma(\Delta_i)} \int_0^{+\infty}\frac{dt_i}{t_i}t^{\Delta_i}\int_0^{+\infty}\frac{dx_0}{x_0^{h+1}}e^{-x_0+\frac{T^2}{x_0}}\\
=&ig\pi^h\prod_{i=1}^3\frac{C_{\Delta_i}}{\Gamma(\Delta_i)} \int_0^{+\infty}\frac{dt_i}{t_i}t^{\Delta_i}e^{T^2}\int_0^{+\infty}\frac{dx_0}{x_0^{h-\frac{\sum_{i=1}^3\Delta_i}{2}+1}}e^{-x_0}\\
=&ig\pi^h\Gamma\left(\frac{\sum_{i=1}^3\Delta_i-d}{2}\right)\prod_{i=1}^3\frac{C_{\Delta_i}}{\Gamma(\Delta_i)} \int_0^{+\infty}\frac{dt_i}{t_i}t_i^{\Delta_i}e^{-t_1t_2P_{12}-t_1t_3P_{13}-t_2t_3P_{23}}\ ,
\ee
where $P_{ij}\equiv -2P_i\cdot P_j$.

For the integration over Schwinger parameters $t_i$, we change the variables $m_{ij}=t_it_j$. Then, 
\be
\label{mij}
\left\langle \mathcal{O}_1(P_1)\mathcal{O}_2(P_2)\mathcal{O}_3(P_3)\right\rangle=&ig\frac{\pi^h}{2}\Gamma\left(\frac{\sum_{i=1}^3\Delta_i-d}{2}\right)\prod_{i=1}^3\frac{C_{\Delta_i}}{\Gamma(\Delta_i)} \prod_{i<j}^3\int_0^{+\infty}\frac{dm_{ij}}{m_{ij}}m_{ij}^{\gamma_{ij}}e^{-m_{ij}P_{ij}}\\
=&ig\frac{\pi^h}{2}\Gamma\left(\frac{\sum_{i=1}^3\Delta_i-d}{2}\right)\prod_{i=1}^3\frac{C_{\Delta_i}}{\Gamma(\Delta_i)}\prod_{i<j}^3\Gamma(\gamma_{ij})P_{ij}^{-\gamma_{ij}}\ ,
\ee
which gives the three-point scalar Mellin amplitude (\ref{3ptscalar}).

Similarly, for the three-point gluon amplitude, we have
\begin{multline}
\left\langle J_1(P_1)J_2(P_2)J_3(P_3)\right\rangle\\
=-igf^{a_1a_2a_3}\int_{\text{AdS}}dX\bigg[ \eta_{A_1A_2} \left(\frac{\partial}{\partial X_{A_3}}\mathcal{E}^{M_2A_2}(P_2,X)\right)\mathcal{E}^{M_2A_2}(P_2,X)\mathcal{E}^{M_3A_3}(P_3,X)\\
-(1\leftrightarrow 2)+\text{cyclic permutations}\bigg]\\
=-igf^{a_1a_2a_3}\Di{3}\Big[2 \eta_{A_1A_2}P_{2,A_3}\prod_{i=1}^3\int_0^{+\infty}\frac{dt_i}{t_i}t^{\Delta_1}t^{\Delta_2+1}t^{\Delta_3}\int_{\text{AdS}}dX e^{2T\cdot X}\\
-(1\leftrightarrow 2)+\text{cyclic permutations}\Big]\\
=\frac{\pi^h}{2}\Gamma\left(\frac{\sum_{i=1}^3\Delta_i-d+1}{2}\right)igf^{a_1a_2a_3} \Di{3}\mathscr{I}_{A_1A_2A_3}\prod_{i<j}^3\Gamma(\gamma_{ij})P_{ij}^{-\gamma_{ij}}\ ,
\end{multline}
from which the Mellin amplitude, (\ref{eq:3J1}), can be read off.
\section{Symanzik's Formula}
\label{Symanzik}
As seen in this paper, the Symanzik's formula is crucial for calculating higher point amplitude from lower point amplitudes. Here we review the derivation of this formula.

First, with Schwinger parameters, the l.h.s. of (\ref{symanzik}) can be written as
\be
\int_{\partial\text{AdS}} dQ\prod_{i=1}^n\Gamma(l_i)\left(-2P_i\cdot Q\right)^{-l_i}=\prod_{i=1}^n\int_0^{+\infty}\frac{dt_i}{t_i}t^{l_i}\int_{\partial\text{AdS}} dQe^{2T\cdot Q}\ ,
\ee
where $T\equiv \sum_{i=1}^nt_iP_i$. Since $t_i$ are positive and $P_i$ are null vectors, $T$ must be time-like. In the rest frame where $T=(T^0,T^\mu)=(|T|,0)$, parametrize the boundary of AdS$_{d+1}$ by
\be
Q=\left(\frac{x^2+1}{2},\frac{x^2-1}{2},x^\mu\right)\ .
\ee
Then,
\be
\int_{\partial\text{AdS}} dQe^{2T\cdot Q}=\frac{\pi^h}{|T|^h}e^{-|T|}\ ,
\ee
where $|T|=\sqrt{-\sum_{i,j=1}^nt_it_jP_i\cdot P_j}$.

Change variables $t_i\to t_i|T|$. Then,
\be
\label{tT}
\prod_{i=1}^n\int_0^{+\infty}\frac{dt_i}{t_i}t^{l_i}\int_{\partial\text{AdS}} dQe^{2T\cdot Q}=\pi^h\int_0^{+\infty}\frac{\det (\partial_{t_i}t_j|T|)}{|T|^n}\frac{|T|^{\sum_{i=1}^nl_i}}{|T|^d}\prod_{i=1}^n\frac{dt_i}{t_i}t^{l_i}e^{-|T|^2}\ .
\ee
The Jacobian is
\be
\det (\partial_{t_i}t_j|T|)=\det (\delta_{ij}|T|-t_j\frac{2\sum_{k=1}^nt_kP_k\cdot P_i}{2|T|})=|T|^n \left(1-\frac{\sum_{k,i=1}^nt_kt_iP_k\cdot P_i}{|T|^2}\right)=2|T|^n\ ,
\ee
where in the second step we have used the formula $\det (\delta_{ij}+A_iB_j)=1+\sum_{i}A_iB_i$. For $\sum_{i=1}^nl_i=d$, (\ref{tT}) becomes
\be
2\pi^h\int_0^{+\infty}\prod_{i=1}^n\frac{dt_i}{t_i}t^{l_i}e^{-|T|^2}=2\pi^h\int_0^{+\infty}\prod_{i=1}^n\frac{dt_i}{t_i}t^{l_i}e^{-\sum_{i<j}^nt_it_jP_{ij}}\ .
\ee

Recall that in (\ref{mij}) we have derived that
\be
2\pi^h\prod_{i=1}^3 \int_0^{+\infty}\frac{dt_i}{t_i}t_i^{\Delta_i}e^{-t_1t_2P_{12}-t_1t_3P_{13}-t_2t_3P_{23}}=\pi^h\prod_{i<j}^3\Gamma(\gamma_{ij})P_{ij}^{-\gamma_{ij}}\ .
 \ee
 More generally,
 \be
2\pi^h\prod_{i=1}^n \int_0^{+\infty}\frac{dt_i}{t_i}t_i^{l_i}e^{-\sum_{i<j}^nt_it_jP_{ij}}=\pi^h\int\prod_{i<j}^n\frac{d\gamma_{ij}}{2\pi i}\Gamma(\gamma_{ij})P_{ij}^{-\gamma_{ij}}\prod_{i=1}^n\delta\left(\sum_{j\neq i}^n\gamma_{ij}-l_i\right)\ ,
 \ee
and thus proves the Symanzik's formula (\ref{symanzik}).
\section{Flat-space limit}
\subsection{Review: flat-space limit of scalar correlators}\label{fllim}
In~\cite{Fitzpatrick:2011ia}, it was shown that in the limit of $\gamma_{ij}\to\infty$, a scalar Mellin amplitude $\mathscr{M}(\Delta_i,\gamma_{ij})$ reproduces the flat-space amplitude $\mathscr{A}(p_i)$. In particular, in the case of massless scalar scattering, the flat-space limit is given by
\begin{equation}
\label{flat}
\begin{split}
\mathscr{M}_n(\Delta_i,\gamma_{ij})\approx \frac{\pi^h}{2}\prod_{i=1}^n\frac{C_{\Delta_i}}{\Gamma(\Delta_i)}\int_0^\infty d\beta\beta^{\frac{1}{2}(\sum_{i=1}^n\Delta_i-d)-1}e^{-\beta}\mathscr{A}_n(p_i\cdot p_j=2\beta\gamma_{ij})\ .
\end{split}
\end{equation}
For example, in $\phi^3$ theory, this relation is simply true for three-point amplitude without even taking the limit (actually the Mellin variables are fixed for this case, so no limit can be taken). To see that, we plug $\mathscr{A}_3=g$ in the r.h.s. of (\ref{flat}) and get
\begin{equation}
\begin{split}
g\frac{\pi^h}{2}\prod_{i=1}^3\frac{C_{\Delta_i}}{\Gamma(\Delta_i)}\int_0^\infty d\beta\beta^{\frac{1}{2}(\sum_{i=1}^3\Delta_i-d)-1}e^{-\beta}=g\frac{\pi^{\frac{d}{2}}}{2}\prod_{i=1}^3\frac{C_{\Delta_i}}{\Gamma(\Delta_i
)}\Gamma\left(\frac{\sum_{i=1}^3\Delta_i-d}{2}\right)\ ,
\end{split}
\end{equation}
which is precisely the scalar three-point Mellin amplitude (\ref{3ptscalar}). 
A slightly nontrivial example is the four-point amplitude. In the $s$-channel with the exchange field of scaling dimension $\Delta$, the Mellin amplitude is~\cite{Paulos:2011ie}
\begin{multline}
\mathscr{M}_{\texttt{Exch}}(\Delta_i)=\frac{\pi^h}{2}\prod_{i=1}^4\frac{C_{\Delta_i}}{\Gamma(\Delta_i)}\sum_{n=0}^\infty\frac{g^2}{4n!\ \Gamma(1+\Delta-h+n)\big(\gamma_{12}-\frac{\Delta_1+\Delta_2-\Delta}{2}+n\big)}\\ \times
\Gamma\left(\frac{\Delta_1+\Delta_2+\Delta-d}{2}\right)\left(1-\frac{\Delta_1+\Delta_2-\Delta}{2}\right)_n\\ \times
\Gamma\left(\frac{\Delta_3+\Delta_4+\Delta-d}{2}\right)\left(1-\frac{\Delta_3+\Delta_4-\Delta}{2}\right)_n\ .
\end{multline}
In the flat-space limit, $\gamma_{12}\to\infty$,
\begin{multline}
\mathscr{M}_{\texttt{Exch}}(\Delta_i)\approx \frac{\pi^h}{2}\prod_{i=1}^4\frac{C_{\Delta_i}}{\Gamma(\Delta_i)}\frac{g^2}{4\gamma_{12}}\sum_{n=0}^\infty\frac{1}{n!\Gamma(1+\Delta-d/2+n)}\\ \times
\left.\left(\frac{\partial^n}{\partial t_1^n}\int_0^\infty d\beta_1\beta_1^{\frac{1}{2}(\Delta_1+\Delta_2+\Delta-d)-1}e^{-\beta_1}t_1^{\frac{1}{2}(\Delta_1+\Delta_2-\Delta)-1}\right)\right|_{t_1=1}\\ \times
\left.\left(\frac{\partial^n}{\partial t_2^n}\int_0^\infty d\beta_2\beta_2^{\frac{1}{2}(\Delta_3+\Delta_4+\Delta-d)-1}e^{-\beta_2}t_2^{\frac{1}{2}(\Delta_3+\Delta_4-\Delta)-1}\right)\right|_{t_2=1}\ .
\end{multline}
Rescaling the integration variables $\beta_{1,2}\to \beta_{1,2}/t_{1,2}$ and using the following identity proven in Appendix C.3 of~\cite{Fitzpatrick:2011ia}, namely
\begin{multline}
\label{c3}
\sum_{n=0}^\infty\frac{1}{n!\Gamma(1+\Delta-d/2+n)}\left.\left(\frac{\partial^n}{\partial t_1^n}\frac{\partial^n}{\partial t_2^n}e^{-\frac{\beta_1}{t_1}-\frac{\beta_2}{t_2}}t_1^{-\Delta+\frac{d}{2}-1}t_2^{-\Delta+\frac{d}{2}-1}\right)\right|_{t_1=t_2=1}
= \\ \beta_1^{\frac{d}{2}-\Delta}e^{-\beta_1}\delta (\beta_1-\beta_2)\ ,
\end{multline}
we get
\begin{equation}
\begin{split}
\mathscr{M}_{\texttt{Exch}}(\Delta_i)\approx&\frac{\pi^h}{2}\prod_{i=1}^4\frac{C_{\Delta_i}}{\Gamma(\Delta_i)}\int_0^\infty d\beta \beta^{\frac{1}{2}(\sum_{i=1}^4\Delta_i-d)-1}\frac{g^2}{4\beta\gamma_{12}}e^{-\beta}\ .
\end{split}
\end{equation}
Note that it is consistent with (\ref{flat}) with $\mathscr{A}_{\texttt{Exch}}(p_i)=g^2(2p_1\cdot p_2)^{-1}$ the flat-space amplitude in the $s$-channel. 
\subsection{Spinning flat space limits}\label{fllimgl}
Now we try to generalize (\ref{flat}) to the vector case. In the previous sections we have already seen the similarity between the gluon amplitude and the flat-space counterpart. In particular, we conjecture the dictionaries between them. So, it is natural to expect that in the flat-space limit Mellin amplitudes reduce to the flat-space amplitudes. 

First, for the three-point gluon amplitude (\ref{eq:3J1}), we have the relation
\begin{multline}
\label{flatv3}
\mathscr{M}_\texttt{3v}^{M_1M_2M_3}=\frac{\pi^h}{2} \left(\Di{3} \right) \int_0^\infty d\beta \beta^{\frac{1}{2}(\gamma_1+\gamma_2+\gamma_3-d)-1}e^{-\beta}\mathscr{A}_{3,A_1A_2A_3}(ik_i\to 2\sqrt{\beta}P_i)\ .
\end{multline}

For the four-point gluon amplitude of the contact diagram, the Mellin amplitude (\ref{eq:4Jcontact}) can also be expressed as
\begin{equation}
\label{fl4con}
\begin{split}
\mathscr{M}^{M_1M_2M_3M_4}_{\texttt{Exch}}=\frac{\pi^h}{2}\left(\Di{4} \right)\int_0^\infty d\beta \beta^{\frac{1}{2}(\sum_{i=0}^4\Delta_i-d)-1}e^{-\beta}\mathscr{A}_{\text{contact},A_1A_2A_3A_4}\ .
\end{split}
\end{equation}

Now we look at the $s$-channel of the vector four-point Mellin amplitude (\ref{eq:M_s}). In the limit of $\gamma_{12}\to\infty$, it becomes
\begin{multline}
\mathscr{M}^{M_1M_2M_3M_4}_{\texttt{Exch}}\approx -g^2\frac{\pi^h}{2}f^{a_1a_2b}f^{a_3a_4b}\left(\Di{4}\right)\sum_{n=0}^{\infty}\frac{\left(\Gamma\left(d-1\right)(\frac{d}{2}-n)_n\right)^2}{4n!\Gamma\left(\frac{d}{2}+n\right)\gamma_{12}}\Big\{X_{12}\Big\}\cdot \Big\{X_{34}\Big\}\ .
\end{multline}
The sum over $n$ can be implemented by using (\ref{c3}) with $\Delta=d-1$. So,
\begin{multline}
\label{sumn}
\sum_{n=0}^{\infty}\frac{\left[\Gamma\left(d-1\right)(\frac{d}{2}-n)_n\right]^2}{n!\Gamma\left(\frac{d}{2}+n\right)}
=\sum_{n=0}^\infty\frac{1}{n!\Gamma(d/2+n)}\left.\left(\frac{\partial^n}{\partial t_1^n}\int_0^\infty d\beta_1\beta_1^{d-2}e^{-\beta_1}t_1^{\frac{d}{2}-1}\right)\right|_{t_1=1} \\\times \left. \left(\frac{\partial^n}{\partial t_2^n}\int_0^\infty d\beta_2\beta_2^{d-2}e^{-\beta_2}t_2^{\frac{d}{2}-1}\right)\right|_{t_2=1}
=\int_0^\infty d\beta\beta^{\frac{3d-4}{2}-1}e^{-\beta}\ .
\end{multline}
Therefore,
\begin{equation}
\label{fl4s}
\begin{split}
\mathscr{M}^{M_1M_2M_3M_4}_{\texttt{Exch}}\approx &-g^2\frac{\pi^h}{2}f^{a_1a_2b}f^{a_3a_4b}\left(\Di{4}\right)\int_0^\infty d\beta\beta^{\frac{3d-4}{2}-1}e^{-\beta}\frac{\mathscr{I}_{A_1A_2A_3A_4}}{4\gamma_{12}}\\
=&\frac{\pi^h}{2}\left(\Di{4}\right)\int_0^\infty d\beta\beta^{\frac{3d-4}{2}-1}e^{-\beta}\\
&\times\mathscr{A}_{s-\text{channel},A_1A_2A_3A_4}\left(\frac{i}{k_1\cdot k_2}\to \frac{1}{2\beta\gamma_{12}},ik_i\to 2\sqrt{\beta} P_i\right)\ .
\end{split}
\end{equation}

Combining (\ref{flatv3}), (\ref{fl4con}) and (\ref{fl4s}), we can summarize the flat-space limit for gluon Mellin amplitudes as follows. Namely, as $\gamma_{ij}\to \infty$,
\begin{equation}
\label{Mnflat}
\begin{split}
\mathscr{M}_n^{M_1\cdots M_n}\approx &\frac{\pi^h}{2}\left(\Di{n}\right)\int_0^\infty d\beta\beta^{\frac{\sum_{i=1}^n\Delta_i-d}{2}-1}e^{-\beta} \\
&\times\mathscr{A}_{n,A_1A_2\cdots A_n} \bigg(\frac{i}{\big(\sum_i k_i\big)^2}\to \frac{1}{4\beta\sum_{i<j}\gamma_{ij}},ik_i\to 2\sqrt{\beta} P_i\bigg)\ .
\end{split}
\end{equation}

Now we show a partial proof of (\ref{Mnflat}) for the case where the $(n+1)$-point amplitude can be factorized into an $n$-point amplitude and a three-point amplitude, as in (\ref{nJchannel}) (also see Figure \ref{nchannel}). First, if the $n$-point amplitude satisfies (\ref{Mnflat}) in the flat-space limit, we can write
\begin{multline}
\tilde{\mathscr{M}}_{n,A_1A_2\cdots A_n}(P_1,P_2,\cdots,P_{n-1},\gamma_{ij})\approx \int_0^\infty d\beta\beta^{\frac{(n-1)d-n}{2}-1}e^{-\beta} \\
\times\mathscr{A}_{n,A_1A_2\cdots A_n} \bigg(\frac{i}{\big(\sum_i k_i\big)^2}\to \frac{1}{4\beta\sum_{i<j}\gamma_{ij}},ik_i\to 2\sqrt{\beta} P_i\bigg)\ .
\end{multline}
By plugging it in (\ref{nM}), we get
\begin{multline}
\label{nMflat}
\mathscr{M}_{n+1}^{M_1M_2\cdots M_{n+1}}\approx \frac{\pi^h}{2} \left(\Di{n+1}\right) igf^{a_na_{n+1}b}\Big\{X_{n(n+1)}^M\Big\}\sum_{m=0}^\infty\frac{\V_3^{m,0,0}}{4\Gamma\left(\frac{d}{2}+m\right)\gamma_{n(n+1)}}\\
\times\sum_{\sum_{r<s}^{n-1}n_{rs}=m}\prod_{r<s}^{n-1}\frac{(\gamma_{rs})_{n_{rs}}}{n_{rs}!} \int_0^\infty d\beta\beta^{\frac{(n-1)d-n}{2}-1}e^{-\beta} \\
\times\mathscr{A}_{A_1A_2\cdots A_{n-1}M}^{a_1a_2\cdots a_{n-1}b} \bigg(\frac{i}{\big(\sum_i k_i\big)^2}\to \frac{1}{4\beta\sum_{i<j}(\gamma_{ij}+n_{ij})},ik_i\to 2\sqrt{\beta} P_i\bigg)\ .
\end{multline}
For the sum over $n_{ij}$, we can repeatedly use
\be
\sum_{\sum_{i<j}n_{ij}\text{ fixed}}\prod_{i<j}\frac{(\gamma_{ij})_{n_{ij}}}{n_{ij}!}\frac{1}{\sum_{i<j}(\gamma_{ij}+n_{ij})}= &\frac{(\sum_{i<j}\gamma_{ij})_{\sum_{i<j}n_{ij}}}{(\sum_{i<j}n_{ij})!}\frac{1}{\sum_{i<j}(\gamma_{ij}+n_{ij})}\\
\approx&\frac{(\sum_{i<j}\gamma_{ij})_{\sum_{i<j}n_{ij}}}{(\sum_{i<j}n_{ij})!}\frac{1}{\sum_{i<j}(\gamma_{ij}+n_{ij})-1}\\
=&\frac{(\sum_{i<j}\gamma_{ij}-1)_{\sum_{i<j}n_{ij}}}{(\sum_{i<j}n_{ij})!}\frac{1}{\sum_{i<j}\gamma_{ij}-1}\\
\approx&\frac{(\sum_{i<j}\gamma_{ij}-1)_{\sum_{i<j}n_{ij}}}{(\sum_{i<j}n_{ij})!}\frac{1}{\sum_{i<j}\gamma_{ij}}\ ,
\ee
for each propagator term in (\ref{nMflat}). Therefore, with totally number of propagators $N_p$, we have
\begin{multline}
\sum_{\sum_{r<s}^{n-1}n_{rs}=m}\prod_{r<s}^{n-1}\frac{(\gamma_{rs})_{n_{rs}}}{n_{rs}!}\prod_{\text{propagators}}\frac{1}{\sum_{i<j}(\gamma_{ij}+n_{ij})}\\
\approx \frac{\big(\sum_{r<s}^{n-1}\gamma_{rs}-N_p\big)_m}{m!}\prod_{\text{propagators}}\frac{1}{\sum_{i<j}\gamma_{ij}}\ .
\end{multline}
Here, the sum of $\gamma_{rs}$ can be evaluated at the pole
\be
\sum_{i<j}^{n-1}\gamma_{ij}=\frac{(n-2)(d-1)-\sum_{i=1}^{n-1}\delta_i}{2}-m\ .
\ee
Thus, (\ref{nMflat}) becomes
\begin{multline}
\label{nMflat1}
\mathscr{M}^{M_1M_2\cdots M_{n+1}}_{n+1}\approx \frac{\pi^h}{2} \left(\Di{n+1}\right)igf^{a_na_{n+1}b}\Big\{X_{n(n+1)}^M\Big\}
\frac{1}{4\gamma_{n(n+1)}}\sum_{m=0}^\infty\frac{1}{m!\Gamma\left(\frac{d}{2}+m\right)}\\
\times\left.\left(\frac{\partial^m}{\partial t_1^m}\int_0^\infty d\beta_1\beta_1^{d-2}e^{-\beta_1}t_1^{\frac{d}{2}-1}\right)\right|_{t_1=1}\bigg(\frac{\partial^m}{\partial t_2^m}\int_0^\infty d\beta_2\beta_2^{\frac{(n-1)d-n}{2}-1}e^{-\beta_2} \\
 \mathscr{A}_{A_1A_2\cdots A_{n-1}M}^{a_1a_2\cdots a_{n-1}b} \Big(\frac{i}{\big(\sum_i k_i\big)^2}\to \frac{1}{4\beta_2\sum_{i<j}\gamma_{ij}},ik_i\to 2\sqrt{\beta_2} P_i\Big)t_2^{\frac{(n-2)(d-1)-\sum_{i=1}^{n-1}\delta_i}{2}-N_p-1}\bigg)\bigg|_{t_2=1}\ .
\end{multline}
Rescaling $\beta_1\to\beta_1/t_1$, $\beta_2\to\beta_2/t_2$ in (\ref{nMflat1}) and using (\ref{c3}) with $\Delta=d-1$, we finally arrive at the formula (\ref{Mnflat}) for the $(n+1)$-point Mellin amplitude $\mathscr{M}_{n+1}$.

\bibliographystyle{utphys}
\bibliography{paper}{}

\end{document}